\documentclass[fleqn,usenatbib,useAMS]{mnras}

\DeclareRobustCommand{\VAN}[3]{#2}
\let\VANthebibliography\thebibliography
\def\thebibliography{\DeclareRobustCommand{\VAN}[3]{##3}\VANthebibliography}

\usepackage{amsmath,amssymb}
\usepackage[normalem]{ulem}
\usepackage{textcomp}
\usepackage{bm}
\usepackage{graphicx}
\usepackage{psfrag}
\usepackage[usenames,dvipsnames]{xcolor}
\usepackage[utf8]{inputenc}
\usepackage{multirow}
\usepackage{tabularx}
\usepackage[makeroom]{cancel}
\usepackage{mathtools}
\usepackage{cuted}
\usepackage{comment}
\usepackage[T1]{fontenc}
\usepackage{makecell}
\usepackage{hyperref}
\usepackage{csquotes}
\allowdisplaybreaks[4]
\graphicspath{{./figures/}}

\newcommand{\msun}{{\rm M}_{\odot}}

\newcolumntype{C}[1]{>{\centering\arraybackslash}p{#1}}
\numberwithin{equation}{section}

\title[Cosmology with EMRIs]{
Gravitational wave cosmology with extreme mass-ratio inspirals
}

\author[D.~Laghi et al.]{Danny Laghi$^{1,2,3}$\thanks{e-mail: danny.laghi@l2it.in2p3.it}, 
Nicola Tamanini$^{3,4}$, 
Walter Del Pozzo$^{1,2}$,
Alberto Sesana$^{5,6}$, 
\newauthor
Jonathan Gair$^{4}$, 
Stanislav Babak$^{7}$,
David Izquierdo\,-Villalba$^{5,6}$
\\
$^{1}$ Dipartimento di Fisica ``Enrico Fermi'', Universit\`a di Pisa, Largo Pontecorvo 3, I-56127 Pisa, Italy\\
$^{2}$ INFN, Sezione di Pisa, Largo Pontecorvo 3, I-56127 Pisa, Italy\\
$^{3}$ Laboratoire des 2 Infinis - Toulouse (L2IT-IN2P3), Universit\'e de Toulouse, CNRS, UPS, F-31062 Toulouse Cedex 9, France\\
$^{4}$ Max-Planck-Institut f\"ur Gravitationsphysik, Albert-Einstein-Institut, Am M\"uhlenberg 1, 14476 Potsdam-Golm, Germany\\
$^{5}$ Dipartimento di Fisica ``G. Occhialini'', Universit\`a di Milano - Bicocca, Piazza della Scienza 3, 20126 Milano, Italy\\
$^{6}$ INFN, Sezione di Milano-Bicocca, Piazza della Scienza 3, 20126 Milano, Italy\\
$^{7}$ Laboratoire Astroparticule et Cosmologie, CNRS, Université de Paris, 10 rue Alice Domon et Léonie Duquet, 75013 Paris, France
}

\date{Accepted: 2021 September 11. Revised: 2021 September 10. Received: 2021 February 5}

\pubyear{2021}

\begin{document}

\label{firstpage}
\pagerange{\pageref{firstpage}--\pageref{lastpage}}
\maketitle

\begin{abstract}
The Laser Interferometer Space Antenna (LISA) will open the mHz frequency window of the gravitational wave (GW) landscape.
Among all the new GW sources expected to emit in this frequency band, extreme mass-ratio inspirals (EMRIs) constitute a unique laboratory for astrophysics and fundamental physics.
Here we show that EMRIs can also be used to extract relevant cosmological information, complementary to both electromagnetic (EM) and other GW observations.
By using the loudest EMRIs (SNR$>$100) detected by LISA as dark standard sirens, statistically matching their sky localisation region with mock galaxy catalogs, we find that constraints on $H_0$ can reach $\sim$1.1\% ($\sim$3.6\%) accuracy, at the 90\% credible level, in our best (worst) case scenario.
By considering a dynamical dark energy (DE) cosmological model, with $\Lambda$CDM parameters fixed by other observations, we further show that in our best (worst) case scenario $\sim$5.9\% ($\sim$12.3\%) relative uncertainties at the 90\% credible level can be obtained on $w_0$, the DE equation of state parameter.
Besides being relevant in their own right, EMRI measurements will be affected by different systematics compared to both EM and ground-based GW observations. Cross validation with complementary cosmological measurements will therefore be of paramount importance, especially if convincing evidence of physics beyond $\Lambda$CDM emerges from future observations. 
\end{abstract}

\maketitle

\begin{keywords}
black hole physics -- gravitational waves -- cosmological parameters 
\end{keywords}


\section{Introduction} 
\label{sec:introduction}

The first direct detection of gravitational waves (GWs) in 2015~\citep{Abbott:2016blz} ended a long experimental quest and opened a new observational window onto the Universe.
Since then, the three ground-based interferometers currently operated by the LIGO-Virgo Collaboration, have announced a total of 50 candidate GW events emitted from binary black hole (BBH) mergers, binary neutron star (BNS) mergers, and possible binary neutron star-black hole mergers~\citep{LIGOScientific:2018mvr,Abbott:2020uma,Abbott2020:gwtc2}.
These observations have shed new light on the physics of compact objects \citep{LIGOScientific:2018mvr,LIGOScientific:2018jsj} and have allowed tests of General Relativity in the dynamical strong field regime for the first time \citep{TheLIGOScientific:2016src,Abbott:2018lct,LIGOScientific:2019fpa, Abbott2020:tgr2}.
With observational runs currently ongoing and improvements in the sensitivity of the interferometers planned for the forthcoming years, many more detections are expected with a progressive increase in the parameter estimation accuracy.
These new observations are quickly consolidating GWs into a new observational science, namely \textit{GW astronomy} \citep{Barack:2018yly}.
Although the direct scope of GW observations consists in recovering the intrinsic and astrophysical properties of GW sources, interesting cosmological information can be extracted from the data as well, especially if associated electromagnetic (EM) counterparts can be identified.
Cosmological inference can thus be considered a subpart of GW astronomy, which we can call \textit{GW cosmology}.

\subsection{Gravitational-wave cosmology} 
\label{sub:gw_cosmology}

The poster example of how GW observations can be used to infer cosmological constraints is given by GW170817, the first BNS merger ever detected \citep{TheLIGOScientific:2017qsa}.
The observation of an EM counterpart to the GW signal \citep{Monitor:2017mdv} allowed the identification of the host galaxy of the source, and thus the use of the event as a cosmic distance indicator. In this context such GW sources are usually referred to as \textit{standard sirens} \citep{Schutz:1986gp,Holz:2005df,Nissanke:2009kt}.
The simultaneous measurement of both the luminosity distance (from the GW waveform) and the redshift (from EM observations) of a GW source provides data points to fit the so-called \textit{distance-redshift relation} \citep{Peebles1993,Weinberg2008cosmology}, which links the luminosity distance to the redshift of each point in the Universe and is a function of the cosmological parameters characterizing the cosmic background expansion.
At low redshift this relation becomes the \textit{Hubble law}, that only depends on the Hubble constant $H_0$.
Given the low redshift ($z\simeq0.01$) of GW170817, this event provided constraints on $H_0$ only. The results are in general agreement, though not competitive, with previous measurements \citep{Abbott:2017xzu}.
Future observations of similar events will reduce the uncertainty on $H_0$~\citep{Dalal:2006qt,Nissanke:2013fka,Chen:2017rfc}, and possibly will help solving the tension on its measured value between local and CMB observations (e.g.,~\citealt{Ade:2015xua,Aghanim:2018eyx,Riess:2016jrr,Riess:2019cxk,Mortsell:2018mfj,Feeney:2018mkj}).

An EM counterpart is not strictly necessary to use compact binary mergers such as BBHs and BNSs as standard sirens.
By matching the sky localisation region of GW sources -- which can be inferred from the GW measurements -- with galaxy catalogs, one might in fact be able to extract complementary information on the redshift of the sources, without the need of spotting an EM counterpart.
The idea was originally proposed by \citet{Schutz:1986gp} and it has subsequently been used and developed in different analyses~\citep{Holz:2005df,MacLeod:2007jd,DelPozzo:2011yh,Petiteau:2011we,Chen:2017rfc,Gray:2019ksv}.
It has already been tested with real data collected by the LIGO and Virgo detectors~\citep{Fishbach:2018gjp,Soares-Santos:2019irc,Abbott:2019yzh,Palmese:2020aof,Finke:2021aom}, though the constraints obtained so far with this ``statistical'' method are not competitive with the ones derived from GW170817 and its EM counterpart, mainly because of the poor spatial resolution of the current network of ground-based interferometers.
Future observations, taken with an enlarged network of ground-based GW detectors, will allow for better cosmological measurements~\citep{Chen:2017rfc}, mainly thanks to the improved sky localisation accuracy. Other complementary methods, which analogously do not require the identification of an EM counterpart, might yield interesting results as well~\citep{Taylor:2012db,DelPozzo:2015bna,Oguri:2016dgk,Mukherjee:2018ebj,Mukherjee:2019wfw,Mukherjee:2019wcg,Mukherjee:2020hyn,Mukherjee:2020mha,Farr:2019twy,Ezquiaga:2020tns}.
The era of precise cosmological measurements with GWs will however start only with next generation interferometers, such as the \textit{Einstein Telescope} (ET)~\citep{Punturo:2010zz,Maggiore:2019uih,Sathyaprakash:2009xt,Belgacem:2019tbw} and the \textit{Cosmic Explorer}~\citep{Abbott:2017CE, reitze:2019program_CE, reitze:2019cosmic_CE}
on the Earth, or \textit{TianQin}~\citep{Mei:2020}, \textit{Taiji}~\citep{taiji:2020102918}, and the \textit{Laser Interferometer Space Antenna} (LISA)~\citep{Audley:2017drz} in space. The latter instrument is the focus of the present investigation.
In what follows, we will briefly introduce LISA and review previous studies of LISA's capability to do cosmological analyses using standard sirens.
More details on how to extract cosmology from GWs by statistically matching with galaxy catalogs will be given in Sec.~\ref{sec:gw_cosmology_without_em_counterparts}.

\subsection{Cosmology with LISA} 
\label{sub:cosmology_with_lisa}

LISA is a space mission designed to detect GWs.
It has been selected by the European Space Agency (ESA) for the L3 slot of its \textit{Cosmic Vision} program, with launch expected in the early 2030s \citep{Audley:2017drz}.
By using interferometric technology already tested in space \citep{Armano:2016bkm,Armano:2018kix}, with a much larger arm-length baseline ($2.5\times 10^6$ km) compared to present ground-based detectors ($\sim$~few km), LISA aims at measuring GWs in the mHz frequency band, which is expected to be populated by GWs emitted by many different sources.

Expected GW sources include: massive black hole binary (MBHB) mergers \citep{Klein:2015hvg}, with masses ranging from $10^4\, \msun$ to $10^7\, \msun$ and detectable up to redshift $\sim$20; the inspiral of stellar-origin black hole binaries (SOBHBs) \citep{Sesana:2016ljz}, with masses ranging from a few tens up to ${\sim}100\, \msun$, the merger of which will be detectable by ground-based interferometers; extreme mass-ratio inspirals (EMRIs) \citep{Babak:2017tow}, which are BBH systems formed by a massive black hole (MBH) and a stellar-origin black hole (SOBH); Galactic and Local-Group binaries \citep{Breivik:2017jip,Korol:2017qcx,Korol:2018ulo,Lau:2019wzw}, i.e.,~compact stellar binaries in the Milky Way and nearby galaxies; and stochastic backgrounds of GWs \citep{Caprini:2015zlo,Bartolo:2016ami,Caprini:2018mtu}, of both cosmological and astrophysical origin.
The measurement and analysis of all these sources will yield unprecedented astrophysical and fundamental physics information, and will allow us to test General Relativity in as yet unprobed regimes.

The GW sources that LISA will observe at cosmological distances can be used as standard sirens.
These include MBHBs, EMRIs, and SOBHBs.
Unfortunately, only for the first of these types of sources are EM counterparts plausibly expected to be produced and observed by future EM facilities \citep{Tamanini:2016zlh}.
MBHBs are in fact expected to emit a large amount of EM radiation in different bands at merger or during long-lasting ($\sim$~weeks/months) afterglows~\citep[see, e.g.,][]{Palenzuela:2010nf,Dotti:2011um,Giacomazzo:2012iv,Moesta:2011bn}, and possibly even through pre-merger signals \citep{Kocsis:2007yu,OShaughnessy:2011nwl,Kaplan:2011mz,Haiman:2017szj,DalCanton:2019wsr}.
If sufficiently accurate sky localisation can be attained from the GW parameter estimation analysis and if the EM counterpart is sufficiently powerful to be spotted by EM telescopes, then we expect to identify the host galaxy of up to a few LISA MBHB mergers per year \citep{Tamanini:2016zlh,Tamanini:2016uin}.
These golden sources can then be used as high-redshift standard sirens to map the expansion of the Universe up to $z\sim 10$.
Although the low number of expected EM counterparts and the high redshift of MBHB mergers are not ideal to test standard cosmological models such as $\Lambda$CDM or to place constraints on late-time dark energy (DE) \citep{Tamanini:2016zlh,Tamanini:2016uin,Belgacem:2019pkk}, they can efficiently be used to probe deviations from $\Lambda$CDM at earlier cosmological epochs, specifically in the interval $3\lesssim z \lesssim 10$ \citep{Caprini:2016qxs,Cai:2017yww,Belgacem:2019pkk,Speri:2020hwc}.
Standard siren analyses with MBHBs would moreover definitely benefit from a network of space-based detectors, e.g.,~LISA and Taiji, which would greatly improve the sky location accuracy of each MBHBs and thus provide better chances to spot the EM counterpart \citep[see, e.g.,][]{Wang:2020dkc,Yang_2021,2021arXiv210502943S}.

At lower redshift, LISA will provide other GW sources that can be used as standard sirens. SOBHBs will be mainly detected at redshifts $z\lesssim 0.1$, while EMRIs might be observed up to $z\sim4$, with a broad peak around $z\approx 1$.
Unfortunately, the most widely accepted formation channels for these types of sources do not predict associated EM counterparts~\citep[see, e.g.,][]{2002ApJ...572..407B,2007CQGra..24R.113A,2016PhRvD..93h4029R}, implying that statistical matching with galaxy catalogs will be necessary in order to extract cosmological information from them (see however~\citealt{2017ApJ...835..165B,2019ApJ...886L..22W,2019BAAS...51c..10E}, for possible EM counterparts of SOBHBs and EMRIs).
A few investigations have already assessed the potential of LISA to test the Hubble law by statistically matching the sky localisation region of SOBHBs with galaxy catalogs~\citep{Kyutoku:2016zxn,DelPozzo:2017kme}.
It was found that constraints on the Hubble constant can reach at best a few \%, though uncertainties on the expected sensitivity of LISA at high-frequency could well undermine this result~\citep{Moore:2019pke}.
No thorough investigation has so far been performed considering EMRIs as possible standard siren sources.
The only analysis that can be found in the literature~\citep{MacLeod:2007jd} suggests that $\sim$20 EMRIs detected at $z\sim 0.5$ could be used to constrain the Hubble constant at the 1\% level.
However, beside considering an old configuration of LISA, this study was highly idealized: the authors assumed only linear cosmic expansion neglecting the acceleration of the Universe, they employed a simplified statistical framework and did not perform parameter estimation over the GW signals, using approximate relations only.
A complete cosmological investigation with LISA EMRIs is currently missing in the literature.

\subsection{Outline} 
\label{sub:outline_and_overview}

The scope of the present investigation is to provide an in-depth and up-to-date analysis of the prospects of using EMRIs detected by LISA as GW standard sirens.
As already stressed and cited (above), the standard capture scenario to generate EMRIs does not predict observable EM counterparts. 
We thus employ a statistical method,  assigning to each galaxy within the LISA 3D error volume a probability of being the host of the EMRI within a Bayesian framework.
We will start by presenting the statistical methodology that we use to infer cosmological parameters by cross-matching GW sky localisation regions with galaxy catalogs (Sec.~\ref{sec:gw_cosmology_without_em_counterparts}).
We will then describe the catalogs of EMRIs observed by LISA that we will use in our analysis (Sec.~\ref{sec:detecting_emris_with_lisa}), and outline the procedure for matching sky localisations with galaxy catalogs (Sec.~\ref{sec:cross_correlation_with_galaxy_catalogues}).
We will subsequently present the results of our investigation (Sec.~\ref{sec:results}), discuss them (Sec.~\ref{sec:discussion}), and finally we will draw our conclusions (Sec.~\ref{sec:conclusion}).


\section{Gravitational-Wave cosmology without EM counterparts} 
\label{sec:gw_cosmology_without_em_counterparts}

To infer the value of the cosmological parameters, we operate within the framework of Bayesian inference~\citep{jaynes:2003}. The starting point of our analysis is: i) a list of GW EMRI observations $D$, combined with ii) a catalog of galaxies that are potential hosts of each individual EMRI. We will describe the EMRI and galaxy catalogs in Sec.~\ref{sec:detecting_emris_with_lisa} and~\ref{sec:cross_correlation_with_galaxy_catalogues}, respectively. Here we provide the mathematical details of the analysis, given those populations.

Our inference model is constructed starting from~\citet{DelPozzo:2011yh}, with updates from~\citet{DelPozzo:2015bna, DelPozzo:2017kme} to account for the uncertainty on the redshift of potential counterparts to GW events. 
We will, however, use a different notation from both references.
In the Bayesian framework, all knowledge about the cosmological parameters we are interested in, $\Omega \equiv \{ H_0,\Omega_m, w_0,w_a\}$, is summarised by the \emph{posterior} probability distribution,
\begin{equation}\label{eqn:bayes-cosmo}
    p(\Omega\,|\,D\,\mathcal{H}\,I) = p(\Omega\,|\,\mathcal{H}\,I)\,\frac{p(D\,|\,\Omega\,\mathcal{H}\,I)}{p(D\,|\,\mathcal{H}\,I)} \,,
\end{equation}
where $\mathcal{H}$ is the cosmological model, that defines the relation between redshift, distance, and cosmological parameters,
$I$ represents all background information relevant for the inference of $\Omega$, and $D\equiv \{ D_1,\ldots,D_N \}$ is the set of GW observations, with $D_i$ the data from the $i$'th EMRI event. 
In a realistic setting, this data would be the strain time series corresponding to the event observed by the LISA detector, so different EMRI events can overlap in time and therefore share this time series. For the current analysis, we will however represent the observed data instead by $D_i = \{\hat{d}_L,\hat{\theta}_{gw},\hat{\phi}_{gw}\}_i$, that is, point estimates of the event luminosity distance and sky position in ecliptic coordinates, with associated uncertainties, and we assume these estimates for each event are independent.

The terms on the right hand side of Eq.~(\ref{eqn:bayes-cosmo}) are the \emph{prior} probability distribution $p(\Omega\,|\,\mathcal{H}\,I)$, 
the \emph{likelihood} function $p(D\,|\,\Omega\,\mathcal{H}\,I)$, and the \emph{evidence} $p(D\,|\,\mathcal{H}\,I)$. Since we are not interested in doing model selection, at this stage the evidence is considered as a normalization constant for the posterior, so we will only define the prior and the likelihood, eventually renormalizing Eq.~\eqref{eqn:bayes-cosmo} at the end of the computation. 

To start with, we formalize our prior information and cosmological model in the following propositions:
\begin{itemize}
    \item[$I$:] ``A GW event can be hosted by only one galaxy. In general not all the galaxies within the comoving volume are visible. Observed and non-observed galaxies obey the same cosmology. 
    Galaxies are highly clustered with each other.''
    \item[$\mathcal{H}$:] ``The cosmological model obeys a flat ($k=0$) Friedmann-Lema\'itre-Robertson-Walker (FLRW) metric~\citep{Weinberg1972} with negligible energy contribution from radiation ($\Omega_r = 0$),     predicting that the luminosity distance $d_L$ is a function of the redshift $z_{gw}$ and of the cosmological parameters $\Omega$: $d_L = d(\Omega,z_{gw})$~\citep{Peebles1993, 1999astro.ph..5116H, Weinberg2008cosmology}''.
\end{itemize}
With assumption $I$, we are stating that the catalogue of galaxies is representative of the whole galaxy distribution, and it is equivalent to assuming that the EMRI lies in one of the galaxies in the catalogue. 
EMRIs hosted in galaxies not in the catalogue are assumed to be sufficiently close to galaxies in the catalogue that this assumption holds. The likelihood for every event is determined by the same catalogue of galaxies, but for ease of computation we only use a subset of possible galaxies to analyse each event. 
Possible hosts are any galaxies that lie within the 2$\sigma$ credible region for the GW direction and distance, for at least one value of the cosmological parameters within the prior range. This selects a 3D co-moving volume in direction and redshift (referred to as an ``error box'') containing $N_{g,i}$ possible hosts, each with sky position ($\theta_j,\phi_j$) and redshift $z_j$. 

We will further specify our model $\mathcal{H}$ investigating two different \emph{cosmological scenarios}, that will define the subset of $\Omega$ which we will be interested in.
In this work we consider two models:
\begin{enumerate}
\item A $\Lambda$CDM scenario characterized by a parameter space $(h,\Omega_m)$ to be explored ($h \equiv H_0/100$ km$^{-1}$ s Mpc, while $\Omega_{DE}$ is determined through the boundary condition $\Omega_m + \Omega_{DE} = 1$, which holds assuming spatial flatness).
We consider a uniform prior range  $h\in[0.6,0.86]$ and $\Omega_m\in[0.04,0.5]$, with fiducial values dictated by the Millennium run~\citep{2005Natur.435..629S} (i.e.,~$h=0.73$, $\Omega_m=0.25$, $\Omega_{DE}=0.75$~\footnote{Although these values of cosmological parameters are outdated, this is irrelevant for the purpose of testing our statistical methodology for constraining cosmology with standard sirens.}).
\item A dark energy (DE) scenario in which we assume $(h,\Omega_m,\Omega_{DE})$ to be pre-determined by other probes at the values of the Millennium run~\citep{2005Natur.435..629S} and we search on the parameters $(w_0,w_a)$ defining the DE equation of state~\citep{Linder:2003} $w(z) = w_0 + z/(1+z) w_a$. We draw from the uniform prior ranges $w_0\in[-3,-0.3]$, $w_a\in[-1,1]$, with fiducial values $w_0 = -1$ and $w_a = 0$, corresponding to the cosmological constant $\Lambda$. 
\end{enumerate}

We assume each GW event is statistically independent from any other event. 
Astrophysically, every EMRI event occurs independently of the others since we will most likely never observe multiple EMRIs occurring in the same galaxy. 
There is, however, coupling between parameter estimation for the events because they will be overlapping in the LISA data set and hence subject to the same noise fluctuations. 
We expect the observed EMRIs to be essentially orthogonal to each other, i.e., the posteriors for each EMRI occupy a small volume in the parameter space with a very small probability of overlap between them. 
So, mutual independence should be a good approximation. 
Thus, the likelihood function simplifies to the product of the likelihoods for each individual observation,
\begin{equation}
    p(D\,|\,\Omega\,\mathcal{H}\,I) = \prod_{i=1}^N p(D_i\,|\,\Omega\,\mathcal{H}\,I)\,.
\end{equation}
Therefore, we only need to determine how to construct the likelihood for a single GW event, whose main ingredients we introduce in what follows. The relevant quantities for the inference of the cosmological parameters are the GW luminosity distance $d_L$ (which is directly measured) and redshift $z_{gw}$ (which is inferred through the cosmology priors)~\footnote{To ease the notation, when there is no possibility of confusion we will be dropping the EMRI index $i$, as in the quantities $d_L$ and $z_{gw}$.}. Assuming that the correlation between sky localisation and other parameters is negligible, a multiple application of marginalisation and product rule lead to a single-event likelihood written as:
\begin{equation}\label{eqn:single-like_first}
\begin{split}
p(D_i\,|\,\Omega\,\mathcal{H}\,I) \!=\!\!
\int \! &\textup{d}d_L \,\textup{d}z_{gw}\,\,
p(d_L \,|\, z_{gw}\,\Omega\,\mathcal{H}\,I)\,\cdot\\
&p(z_{gw} \,|\, \Omega\,\mathcal{H}\,I)\,
p(D_i \,|\, d_L\,z_{gw}\,\Omega\,\mathcal{H}\,I)\,,
\end{split}
\end{equation}
where the integrals on $d_L$ and $z_{gw}$ go, in principle, from 0 to $\infty$. 

Once we know the redshift $z_{gw}$ along with the values of the cosmological parameters (assuming their priors) and the cosmological model $\mathcal{H}$, 
we have \citep{DelPozzo:2011yh}
\begin{equation}\label{eqn:lk_dL}
    p(d_L\,|\,z_{gw}\,\Omega\,\mathcal{H}\,I) = \delta(d_L - d(\Omega, z_{gw}))\,,
\end{equation}
where under our background information $I$ the function $d(\Omega, z_{gw})$ is given by \citep{1999astro.ph..5116H}:
\begin{equation}\label{eqn:dl_relation}
d(\Omega, z_{gw}) = \frac{c(1+z_{gw})}{H_0} \int_0^{z_{gw}} \!\! \frac{\textup{d}z'}{E(z')} \,,
\end{equation}
with the function $E(z')$ given by:
\begin{equation}\label{eqn:E_z}
E(z') = \sqrt{\Omega_m(1+z')^3 + \Omega_{DE}\,g(z', w_0, w_a)}\,,
\end{equation}
where
\begin{equation}
g(z', w_0, w_a) = (1+z')^{3(1+w_0+w_a)}e^{-3\frac{w_a z'}{1+z'}}\,,
\end{equation}
and as already noted in the definition of $\mathcal{H}$ we neglect the radiation density $\Omega_r$, since $\Omega_r \ll \Omega_m,\Omega_{DE}$. 

Our prior information $I$ prescribes that each GW is hosted by a galaxy; we are assuming that the distribution of $z_{gw}$ is the same as that of the host galaxies. Thus we have
\begin{equation}\label{eqn:prior_zgw}
    p(z_{gw}\,|\,\Omega\,\mathcal{H}\,I)  \propto
    \sum_{j=1}^{N_{g,i}} w_j 
    \exp{\Biggl\{-\frac{1}{2}\biggl(\frac{z_j-z_{gw}}{\sigma_{z_j}}\biggr)^{\!2}\Biggr\}} \,,
\end{equation}
where we have introduced relative weights $w_j$ for each galaxy host (this is different from \citet{MacLeod:2007jd}, where all galaxies within the error-box were assumed to be equally likely hosts). The uncertainty $\sigma_{z_j}$ includes the contribution due to the peculiar velocity of the galaxy $j$, and the sum goes over all the possible galaxy hosts $N_{g,i}$ associated to the $i$-th event.
As in \citet{DelPozzo:2017kme}, we assume a typical redshift error of $\Delta z = 0.0015$, 
corresponding to a rms peculiar velocity of 500 km/s, which is used as the Gaussian standard deviation in the redshift error.
We assign the relative weights $w_j$ to each galaxy from the marginal distribution over the sky position angles computed in each galaxy: the weights $w_j$ account for the inferred LISA sky location of the source.
When computing Eq.~\eqref{eqn:single-like_first}, we should in fact also be integrating over $\theta_{gw}$ and $\phi_{gw}$, and the prior coming from the galaxy catalogue should include delta functions in each term, centered on the angular coordinates of the potential hosts. 
However, if correlations between the LISA measurements of the sky position angles and distance are neglected (which is equivalent to assuming that the joint likelihood on sky location and distance factorises), the integrals on $\theta_{gw}$ and $\phi_{gw}$ can be done directly and return a marginal distribution over $\theta_{gw}$ and $\phi_{gw}$, computed at the sky location of the galaxies. This yields the number $w_j$, as defined by Eq.~\eqref{eq:margweights} and described in detail in Sec.~\ref{sec:cross_correlation_with_galaxy_catalogues}. We note that the product $w_j\,p(D_i \,|\, d_L\,z_{gw}\,\Omega\,\mathcal{H}\,I)$ should equal the full likelihood evaluated at the sky location of the galaxy and therefore, in principle, $w_j$ depends on both $d_L$ and on the observed data, which is not explicit in the preceding equation.
As a final comment, these data-dependent weights still treat all galaxies as \textit{a priori} equally likely hosts of observed EMRIs. It is only the differing sky locations of the galaxies that affect the a-posteriori probability. Weightings can also be used to reflect the relative probability that a galaxy is the GW host, for example based on the galaxy luminosity. The expressions look the same, but the weights are then proportional to the product of the sky-location contribution and the pre-assigned weight. We will not consider other types of galaxy weighting in this analysis.

The remaining term in Eq.~\eqref{eqn:single-like_first} is:
\begin{equation}\label{eqn:single_like_second}
p(D_i\,|\,d_L\,z_{gw}\,\Omega\,\mathcal{H}\,I) = p(D_i\,|\,d_L\,z_{gw}\,\mathcal{H}\,I)\,,
\end{equation}
which is the likelihood for the GW data. As described above, we approximate the GW data as point estimates of the relevant parameters, with Gaussian uncertainties, and so replace this term by a quasi-likelihood specified as a multivariate Gaussian, with covariances estimated from the Fisher Matrix (see Sec.~\ref{sec:detecting_emris_with_lisa}), and also accounting for the weak lensing uncertainty $\sigma_{W\!L,i}$ as modelled in \citet{Tamanini:2016zlh}, see Eq. (7.3) therein (see also \citealt{Hirata:2010ba,Cusin:2020ezb}).

The final expression for the single-event likelihood is:
\begin{equation}
\begin{split}
\label{eqn:single_event_likelihood}
    &p(D_i\,|\,\Omega\,\mathcal{H}\,I) \! = \frac{1}{2\pi} \int   \textup{d}z_{gw,i} 
         \sum_{j=1}^{N_{g,i}} 
    \frac{w_j}{\sigma_{z_j} \sqrt{\sigma_{\hat{d}_{L,i}}^2\!+\sigma_{W\!L,i}^2}} \cdot\\
    &\exp{\Biggl\{\! -\frac{1}{2} \Biggl[ \frac{\bigl( z_j-z_{gw,i} \bigr)^2}{\sigma_{z_j}^2} +\frac{\bigl(\hat{d}_{L,i} - d(\Omega, z_{gw,i})\bigr)^2}{\sigma_{\hat{d}_{L,i}}^2\!+\sigma_{W\!L,i}^2} \Biggr] \Biggr\}}
\,.
\end{split}
\end{equation}
As mentioned above, $(\hat{d}_L,\cos\hat{\theta}_{gw}, \hat{\phi}_{gw})_i$ are the point estimates of the parameters, and the uncertainties, $\sigma_{d_{L,i}}$ etc., are determined from the Fisher matrix evaluated at the true parameter values. We have also assumed that the joint likelihood on sky location and distance factorises as mentioned above.

Until now, we have assumed that all EMRIs that occur during the observation period are included in the analysis. However, in practice our detectors have limited sensitivity and we will only include in the analysis the events that are successfully ``detected'' by our analysis pipelines, that is, those for which some ranking statistic computed from the data is above some predetermined threshold. Whether or not an event is found is a property of the data only, and assuming that the total number of events does not convey any information the selection effects can be accounted for~\citep{Mandel:2019} by replacing $p(D_i\,|\,\Omega\,\mathcal{H}\,I)$ with
\begin{equation}
    \hat{p}(D_i\,|\,\Omega\,\mathcal{H}\,I) = \frac{p(D_i\,|\,\Omega\,\mathcal{H}\,I)}{\int_{D > \mbox{threshold}}\, p(D_i\,|\,\Omega\,\mathcal{H}\,I) \, {\rm d}D},
\end{equation}
where the integral is over all data sets that would give detection statistics above the threshold and hence be included in a cosmological analysis. 
In practice, the detectability of a GW event is primarily determined by its signal-to-noise ratio (SNR), which in turn is primarily determined by luminosity distance. In the quasi-likelihood model used here we could thus use a cut on the observed luminosity distance, $\hat{d}_L > d_{\rm crit}$, as a proxy to represent selection effects. 
As the choice of cosmological parameters is varied, the centres of the Gaussian distributions in the likelihood will move inside and outside the selection cut. Although a small number of the Gaussian distributions span the boundary, the selection function normalisation can be well approximated by the number of host galaxies that remain consistent with being inside the luminosity distance horizon. In our analysis we found that this number changed by very little over the range of cosmological parameters consistent with our priors, and so we approximated this normalisation as a constant. However, we also verified that our results were insensitive to that approximation. 

The fact that the selection function can be ignored can be understood by considering its fractional effect on the posterior distribution, but the essential reason is that the measurements are quite precise and so the selection function does not change very much over the posterior. For low redshift horizons the selection function scales roughly like $h^3$, obtained by integrating a redshift distribution $p(z) \propto z^2$ out to a fixed luminosity distance horizon (see~\citealt{Chen:2017rfc}). The impact of such a correction on the population likelihood can be assessed from the Fisher matrix, which is given by the expected value of the second derivative of the log-likelihood for an observation. As the selection effect enters as a renormalisation, it makes an additive contribution of $3/h^2\sim 6$ to the Fisher matrix, which is a $\sim1\%$ correction to the value without selection effects and hence negligible. The bias due to omitting this factor can also be estimated, from the first derivative of the log-likelihood, and is $\sim0.005$, which is comfortably below the statistical uncertainties. For larger numbers of events the bias could start to become important, but for all cases considered here corrections from the selection effect are sub-leading.

This completes the definition of our inference problem. We will further discuss the limitations and approximations of our framework in Sec.~\ref{sec:detecting_emris_with_lisa},~\ref{sec:conclusion}, and Appendix~\ref{app:appendix}.

We explore the posterior distribution in Eq.~\eqref{eqn:bayes-cosmo} using \texttt{cpnest}~\citep{cpnest}, a parallel nested sampling algorithm implemented in~\texttt{Python}. We make use of the \texttt{LALCosmologyCalculator} library in \texttt{LAL}~\citep{lal} and of \texttt{numpy}~\citep{numpy},
\texttt{scipy}~\citep{scipy}, \texttt{matplotlib}~\citep{matplotlib}, \texttt{cython}~\citep{cython}, and
the plotting utilities in \citet{corner}. The inference code utilised in this paper is publicly available in~\citet{cosmolisa}.


\section{Detecting EMRIs with LISA} 
\label{sec:detecting_emris_with_lisa}

\subsection{Expected properties of the EMRI population}

The catalogs of EMRIs detected by LISA which we use in our work are based on the analysis of~\citet{Babak:2017tow}.
Here we review how these catalogs have been constructed and outline their main properties.
For more information the reader is referred to~\citet{Babak:2017tow} ~\citep[see also][]{Babak:2014kqa,Gair:2017ynp}.

Because of their small mass-ratio, EMRIs present a very slow inspiral, producing many cycles ($10^4$-$10^5$) within LISA's sensitivity band.
For the same reason, the detailed dynamics of these systems strongly differ from equal mass compact binaries.
The motion of the stellar BH can in fact be approximated by geodesics of the massive BH, with small but relevant corrections due to its own \textit{self-force}~\citep[for a review, see, e.g.,][]{Poisson:2011nh}.
Unfortunately ongoing perturbative calculations, which exploit the extreme difference in masses of these systems, are not yet at the level needed to calculate the emitted GW waveform with the accuracy required for LISA observations (i.e.,~by keeping track of the phase over the entire inspiral).
The analysis of~\citet{Babak:2017tow} considered thus an approximate analytical model to estimate the waveform generated by EMRIs, the so-called \textit{analytical kludge} model~\citep{Barack:2003fp}.
This approximation was used to produce waveforms under two different endpoints for the dynamical evolution of the system: until the Schwarzschild last stable orbit (more pessimistic) and until the Kerr last stable orbit (more optimistic).
In our study we will only consider results  obtained with the latter of these assumptions (denoted ``AKK'' in~\citet{Babak:2017tow}).

In order to simulate the response of LISA and estimate the uncertainty on the parameters of the GW waveform generated by an EMRI, the investigation made by~\citet{Babak:2017tow} employed a \textit{Fisher matrix} approach, useful to quickly analyse many different events.
Their results were obtained setting a threshold of SNR=20 for LISA detection and considering a LISA sensitivity curve as specified by the LISA White Paper written in response to ESA's call for the L3 mission slot~\citep{Audley:2017drz}.
The results of this paper will thus be based on that sensitivity curve as well.
We note here that although the laser stability requirement has been relaxed since, this has an impact to the detector performance at $f>0.01$ Hz, which is unlikely to significantly affect EMRIs.

As pointed out by \citet{Babak:2017tow}, the main uncertainties in forecasting how many EMRIs will be detected by LISA are of astrophysical origin.
The expected rate of EMRIs depends in fact on several astrophysical assumptions, including:
\begin{itemize}
    \item the MBH population in the accessible LISA mass range (from $10^4 \msun$ to $10^7 \msun$), the redshift evolution of their mass function, and their spin distribution;
    \item the fraction of MBHs hosted in dense stellar cusps, which constitute the nurseries for the formation of EMRIs;
    \item the EMRI rate per individual MBH, and the mass and eccentricity distribution of the inspiralling compact object.
\end{itemize}
In the analysis of~\citet{Babak:2017tow}, by considering 12 different combinations of prescriptions for the assumptions listed above, the authors produced forecasts for 12 different scenarios. Among all 12 scenarios, the rates for LISA detections using the AKK model always fall in between 10\% to 20\% of the total EMRI population, and in absolute numbers they span a range from one to a few thousands of  detections per year (at SNR>20), with similar properties between the 12 EMRI populations corresponding to each scenario.
In particular, irrespective of the astrophysical scenario, EMRIs detected by LISA will come from MBHs with masses from $3\times 10^4 \msun$ to $3\times 10^6 \msun$, over a redshift range that is broadly peaked between $0.5 < z < 2$, with tails usually reaching $z\sim5$.

In this paper we concentrate on three EMRI models, spanning the bulk of the uncertainty range in the number of EMRI detections, corresponding to models M1, M5, and M6 of~\citet{Babak:2017tow} (cf.~their Table~I):
\begin{itemize}
\item our {\it fiducial} model is based on M1, which depends on rather standard assumptions for the MBH mass function and EMRI properties; 
\item  our {\it pessimistic} model is based on M5, featuring a downturn of the MBH mass function at $M<10^7\msun$, which strongly suppresses the occurrence of EMRIs;
\item  our {\it optimistic} model is based on M6. This is similar to M1, but ignores the effect of cusp erosion following galaxy mergers, which boosts the EMRI rate by a factor of $\sim2$.
\end{itemize}
Although there are more optimistic and pessimistic EMRI models in~\citet{Babak:2017tow}, those are based on rather ad-hoc assumptions, in particular about the relative occurrence of plunges and EMRIs. In fact, when a BH is scattered on an almost radial orbit, chances are that it plunges directly onto the MBH, rather than forming an EMRI. 
N-body simulations of realistic stellar cusps suggest that there are few such direct plunges for each EMRI~\citep{2015ApJ...814...57M}. The models listed above assume a rather conservative ratio of 10 plunges per EMRI. 
We note that, as the number of useful events $N$ increases further, one can reasonably expect that our final results, i.e.,~the constraints that we obtain on the cosmological parameters, scale approximately with $\sqrt{N}$ (cf.~Sec.~\ref{sec:discussion}). 

Each EMRI model predicts different events, thus it can be thought of as a different data set $D$. For each data set, we will adopt the cosmological \emph{inference} models $\mathcal{H}$ detailed in Sec.~\ref{sec:gw_cosmology_without_em_counterparts}.

\subsection{EMRI selection} 
\label{sec:EMRIsel}

Because of the complexity of their waveform, a relatively large SNR is required for EMRI detection. As already noted, the threshold is customarily set to SNR$=20$, which results in several hundreds-to-thousands of EMRIs detected over 10 years for the three models considered here, as reported in the second column of Table~\ref{tab:Nemri}.

Although such an abundance of sources is in principle an asset for our analysis, a number of considerations have to be made when designing a practical implementation of the algorithm. In fact, the speed of computation of Eq.~\eqref{eqn:single_event_likelihood}
strongly depends on the dimension of the EMRI catalog. This imposes a limitation on the number of events that we can analyse in a reasonable time. At the current state, catalogs of hundreds of EMRIs are prohibitive, as they require $\mathcal{O}(\text{months})$ to be analysed. This is due to two main reasons: i) the increase in the number of single-event likelihoods that have to be computed and ii) the large number of hosts $N_{g,i}$ that some events will have, resulting in a significant slowdown of each likelihood evaluation. Importantly, since the best-localised events are also the most informative, it seems reasonable to exclude from the analysis those events that are not well-localised, since they would not significantly improve our results.

\begin{figure}
\includegraphics[width=0.45\textwidth]{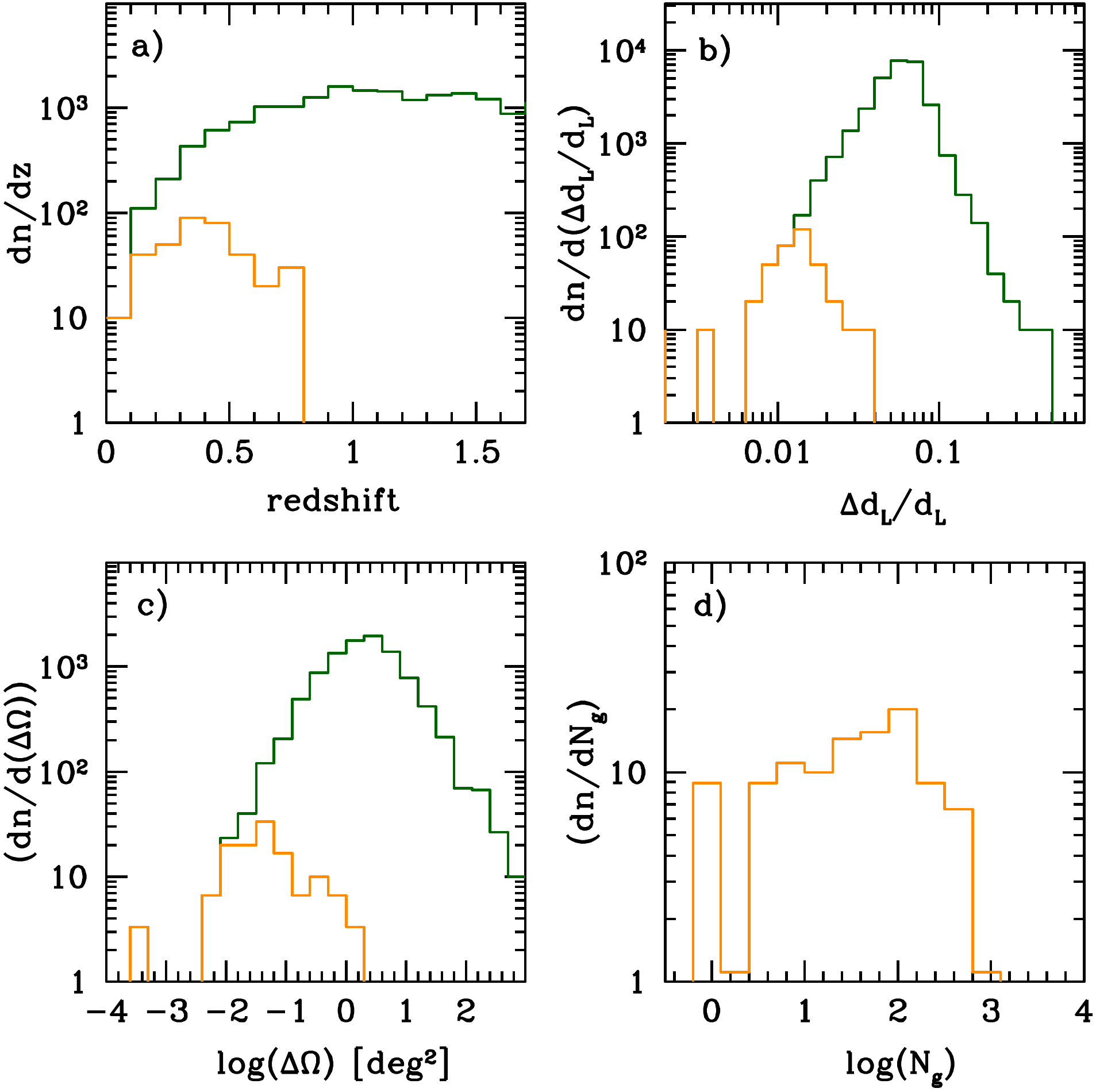}
\caption{Relevant properties of the events used in the analysis, taken from model M1. Green histograms in panels a), b), and c) show the redshift, distance error, and sky location error distributions for the entire population of EMRIs detected by LISA in 10 years of operations. 
For the purpose of our analysis we select only systems with SNR$>$100. Those are shown in panels a), b) and c) as orange histograms. For these latter systems, panel d) shows the distribution of the number of candidate hosts within the 3D 2$\sigma$ error volume, averaged over the three realisations of M1 (see section \ref{sec:cross_correlation_with_galaxy_catalogues} for details).
}
\label{fig:events}
\end{figure}

Point ii) suggests some useful guidelines for devising an EMRI selection strategy. As shown for our fiducial model M1 by the green histograms in Fig.~\ref{fig:events}, the majority of EMRIs will be observed at $z>1$. As $z$ increases, so do the errors in the estimate of their parameters  and the associated LISA 3D localisation volume $\Delta{V}$. 
When $\Delta{V}$ gets too large, the number of candidate galaxy hosts within it, $N_{g,i}$, increases dramatically, washing out the information enclosed in their clustering properties, which is of paramount importance for the success of our technique.

This suggests two approaches to the selection of the events.
One possible strategy, that we indicate as our first selection procedure, is to pre-select EMRIs with a good LISA 3D sky location estimate up to some maximum redshift. Following this route, we select only EMRIs with sky localisation error $\Delta \Omega=2\pi \sin\theta_{gw} \sqrt{\Delta \theta_{gw} \Delta \phi_{gw} - (\Sigma^{\theta_{gw}\phi_{gw}})^2}<2$\,deg$^2$, being $\theta_{gw}$ and $\phi_{gw}$ the latitude and longitude of the source in ecliptic coordinates and $\Sigma^{\theta_{gw}\phi_{gw}}$ their correlation extracted by the  3D marginalised correlation matrix of the Fisher analysis.
We further require a relative error in the luminosity distance determination: $\Delta{d_L}/d_L<0.1$.
Moreover, an additional constraint stems from the fact that, in order to keep the number of candidate hosts within a maximum of a few thousands, 
we use the mock Millennium sky to select galaxy hosts up to $z=1$. 
This implies that we keep only those events for which the furthest candidate host compatible with the adopted priors in the cosmological parameters is at $z<1$. 
In fact, the cut in our galaxy catalogue at $z=1$ implies that we discard some of the events at $z\gtrsim0.7$ for which the furthest candidate compatible with the range of cosmological priors lies at $z>1$. We note that this cut depends on the adopted cosmological model, and yields a slightly different selection for the $\Lambda$CDM and the DE cases. The number of events selected according to this procedure for the $\Lambda$CDM and DE models are given in the third and fourth columns of Tab.~\ref{tab:Nemri}.

Although legitimate, this selection procedure is rather contrived, as it combines a number of ad hoc choices in the selection process. 
A more straightforward alternative 
is to observationally select events by simply imposing an SNR threshold. 
This is illustrated for our fiducial model M1 by the orange histograms in Fig.~\ref{fig:events}.
In fact, high-SNR events are expected to be those yielding a better determination of the EMRI parameters (distance and sky location) and are also preferentially at relatively low redshifts: this is evident looking at panels a), b), and c). 
This combination of properties automatically limits the number of candidate hosts -- see panel d) -- making them the most informative sources for our analysis. 
We found empirically that imposing a LISA SNR threshold SNR$>$100 returns only events with $N_g\lesssim 1000$ out to $z\approx 0.75$, as shown in panel a). 
The imposition of an SNR threshold, together with the cut at $z=1$ dictated by the galaxy catalog that is subsequently used (which is sensitive to the assumed cosmological model),
yields the number of events reported in the fifth and sixth columns of Table~\ref{tab:Nemri}.
The first selection procedure produces a larger number of EMRIs~(see the third and fourth columns of Tab.~\ref{tab:Nemri}), among which we found several events with relatively low SNR and $N_g>1000$. 
Tests performed on these samples produced reasonable estimates of the cosmological parameters, but also displayed some undesired behaviour related to the limitations of our likelihood implementation, which became particularly evident when the majority of the selected events had low SNR. These aspects will be further discussed later on and in particular in Appendix~\ref{sec:app_low_snr}. 
Therefore, unless otherwise stated, results reported in this work follow the SNR$>$100 selection criterion.

\begin{table}
\centering
\begin{tabular}{c|c|c|c|c|c} 
\hline
\hline
 & \multirowcell{3}{
      detected \\
      (SNR$>$20)
} &  \multicolumn{2}{c|}{$\Delta{d_L}/d_L<0.1$}&  \multicolumn{2}{c}{\multirow{2}{*}{SNR$>$100}}\\
MODEL &  & \multicolumn{2}{c|}{$\&\,\Delta{\Omega}<2\,$deg$^2$}  & \multicolumn{2}{c}{}\\
\cline{3-6}
 &  & $\Lambda CDM$ & $DE$ & $\Lambda CDM$ & $DE$ \\
\hline
\hline
{M1\,({\it fid})} & {2941} & 180 & 202 & 32 & 33\\
{M5\,({\it pess})} & {472} & 20 & 21 &  6 & 6\\
{M6\,({\it opt})} & {4788} & 260 & 281 & 72 & 73\\
\hline
\hline
\end{tabular}
\caption{
Number $N$ of EMRIs observed by LISA in 10 years of operation. For each model (first column) we report the total number of detections (second column), those additionally localized within $\Delta{d_L}/d_L<0.1\,\, \&\,\, \Delta{\Omega}<2$deg$^2$, referred in the text as first selection procedure, condition that depends on the chosen cosmological model (third and fourth columns), and those with SNR$>100$ (fifth and sixth columns). In both selection procedures we require no potential host galaxy above $z=1$.
\label{tab:coeff_bckgr}
}
\label{tab:Nemri}
\end{table}

\subsection{Caveats and limitations}

Selecting only high-SNR events allows us to avoid some limitations and complications related to a number of features that are not (as of yet) modeled in our pipeline. 

First, we do not fold into our analysis the possible incompleteness of the galaxy catalogs. As we will see in the next section, galaxy catalogs are constructed starting from the flux-limited mock sky realizations of~\citet{2012MNRAS.421.2904H} based on the Millennium simulation~\citep{2005Natur.435..629S}. 
As reported in our prior information $I$, we are going to assume that each GW event is hosted by one of the galaxies within the reconstructed co-moving volume, regardless of whether the galaxy is luminous enough to be listed in the mock catalog or not. 
We assume that EMRIs are hosted by galaxies with stellar mass $M_*>10^{10}\msun$, for which we found that the Millennium sky maps are complete out to $z\approx0.5$. 
The large majority of events we select are at $z<0.5$ (see Fig.~\ref{fig:events}).
Thus, the mass threshold imposed in constructing our catalogs
should not yield a relevant Malmquist bias. 
This is further discussed in Appendix~\ref{sec:app_gal_mass_thr}.

Another possible source of bias may be given by evolution effects, i.e, the fact that the host galaxy population evolves within the allowed redshift range for the host. Also in this case, the high-SNR cut preferentially selects low-$z$ events with precise distance measurement, thus limiting the redshift of candidate hosts within a window that is narrow enough that evolution effects can be neglected.

Last, this selection leaves out events with a large number of candidate hosts, that are not expected to make significant contributions to the inference. 
When testing our pipeline, we found that less informative, low SNR events not only do not add much information to the inference, but tend to produce a bias in the estimate of cosmological parameters, especially those that are weakly constrained. 
More details about this bias are given in App.~\ref{sec:app_low_snr}. Its origin is still under investigation and will be a subject of future work.


\section{Statistical matching with galaxy catalogues} 
\label{sec:cross_correlation_with_galaxy_catalogues}

\begin{figure*}                        
\includegraphics[width=0.90\textwidth]{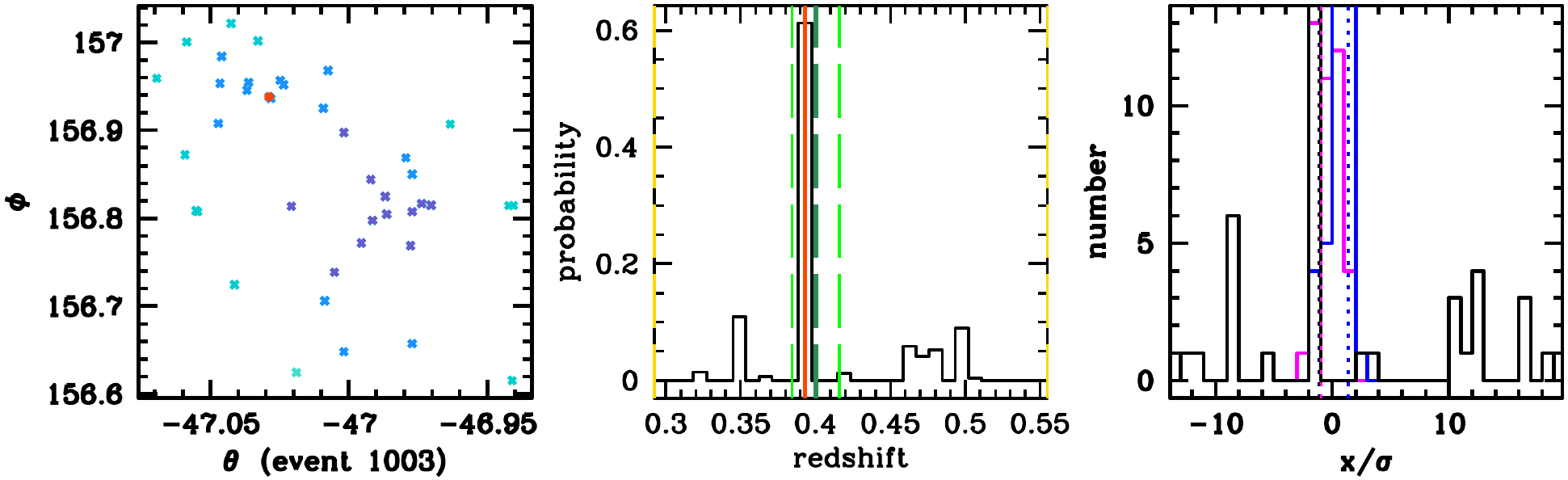}\\
\includegraphics[width=0.90\textwidth]{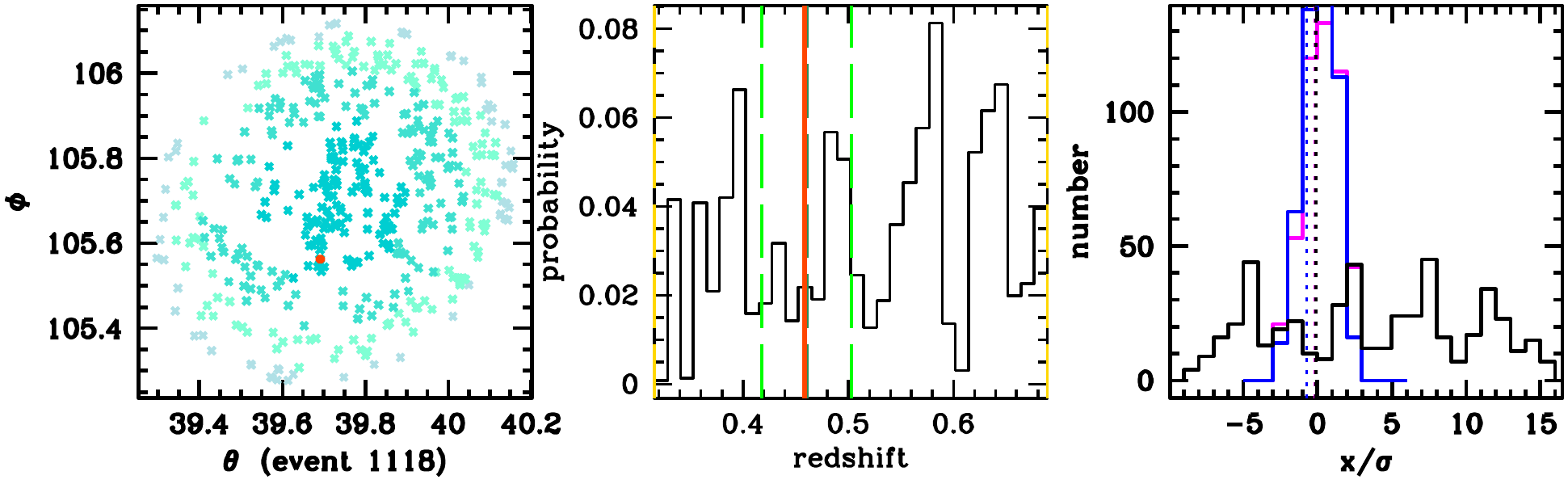}\\
\includegraphics[width=0.90\textwidth]{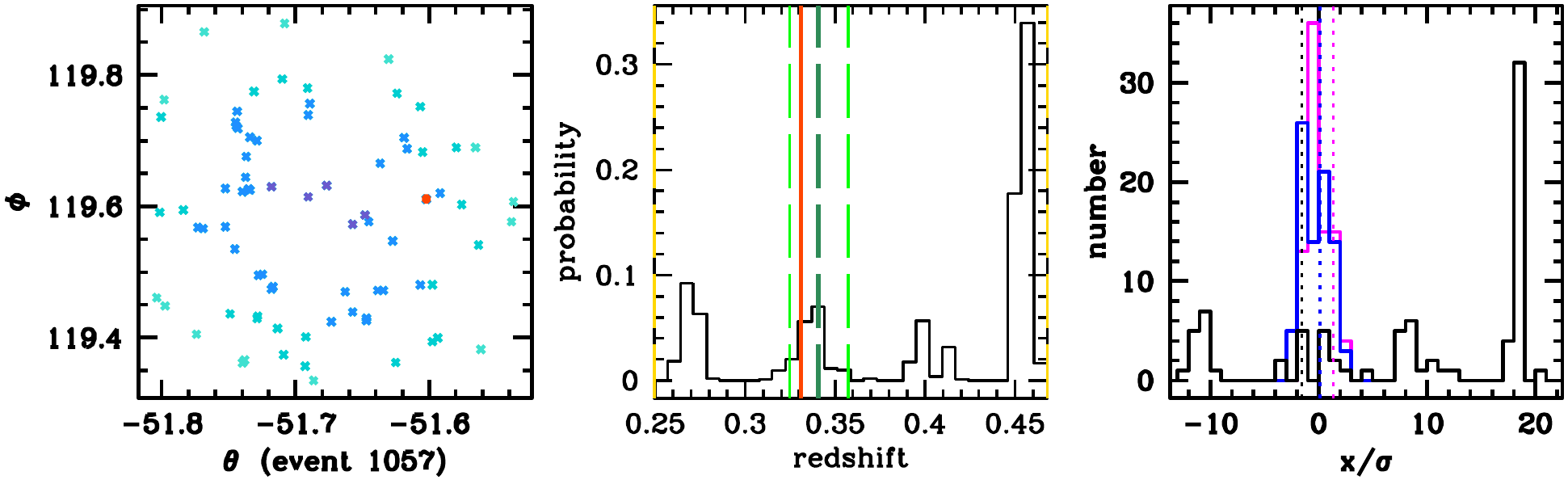}\\
\caption{
Each row of three panels represents the properties of the galaxy host candidates for a selected event extracted from model M1\_2.
The left panels show a scatter plot of candidate hosts in the $\theta-\phi$ plane; the shades of blue, from lighter to darker, are in order of increasing probability of hosting the EMRI, according to the LISA sky localisation in the celestial sphere. 
The red dot is the true host, which is generally not in the center of the sky localisation area. The middle panels show the redshift probability distributions of the candidates hosts (black histograms), i.e. the distribution of galaxies, each weighted by its own probability of being the host according to its sky location.
The dark-green vertical line is the fiducial redshift measured by LISA and the two light-green lines bracket the 2$\sigma$ redshift interval implied by the LISA measurement error in the luminosity distance alone, assuming the true cosmology. The vertical yellow lines at the edges bracket the redshift interval uncertainty when the cosmological parameters are allowed to vary within the assumed priors (here $\Lambda$CDM is assumed). The redshift of the true host is marked by the vertical red line. 
Finally, the right panels represent the deviation of the $x$ quantities, $\theta$ (magenta), $\phi$ (blue), and $d_L$ (black), from the best measured value $x_t$ normalized to the respective LISA measurement errors. The vertical dotted lines are the values of the true hosts. Note that the $d_L$ of candidate galaxies can deviate from the best measured value by many $\sigma$, as measured by LISA, due to the prior uncertainties on the cosmological parameters.}
\label{fig:galaxy}
\end{figure*}

Having selected a sample of ``useful'' EMRIs, the next step is to simulate realistic distributions for candidate host galaxies. To this end, we closely follow the procedure outlined in~\citet{DelPozzo:2017kme}, with some improvements, as we now summarize.

We use the simulated observed full sky available within the \citet{VMD}. The sky is built from the Millennium run~\citep{2005Natur.435..629S}, so that, critically for this work, the simulation reproduces the observed clustering properties of galaxies. The full sky map is constructed following the procedure outlined in \citet{2012MNRAS.421.2904H} for the ``pencil beam'' fields. The only difference is the much shallower depth of the map, which is limited to galaxies with AB magnitude $i<21.0$. A comparison of the galaxy mass functions between the full sky and the much deeper pencil beams shows that the former is completed down to $M_\star=10^{10}\msun$ only at $z\lesssim 0.5$. We consider only galaxies with $M_\star>10^{10}\msun$ as potential hosts. We show in Appendix~\ref{sec:app_gal_mass_thr} that this arbitrary choice does not affect our results significantly.

The catalog, that we use up to $z=1$,
lists a number of galaxy properties, including mass, observed redshift $z_{\rm obs}$ and cosmological redshift $z_{\rm cos}$, which differ due to the extra redshift contribution imprinted on the former by galaxy peculiar velocities, $\Delta{z}_{v_p} \equiv z_{\rm obs} - z_{\rm cos} = (1+z_{\rm cos})v_p/c \,\,({v_p} \ll c)$, $v_p$ being the peculiar velocity.

The simulation assumes a $\Lambda$CDM cosmology with $h=0.73,\,\Omega_m=0.25,\,\Omega_\Lambda=0.75$, which will thus correspond to the fiducial values for our analysis as well.
Although those parameters are outdated~\citep{Ade:2015xua}, this is irrelevant for the illustrative purpose of our analysis. 

The LISA 3D error volume is approximated by a multivariate Gaussian distribution:
\begin{equation}
  p(D_i | {\bf \Theta})=\frac{1}{\sqrt{(2\pi)^3|{\bf \Sigma|}}} \exp\left\{\!-\dfrac{1}{2}{\bf \tilde{\Theta}}_i^{T}{\bf \Sigma}^{-1}{\bf \tilde{\Theta}}_i\right\} \text{,}
\label{eq:psky}
\end{equation}
where ${\bf \tilde{\Theta}}_i \equiv {\bf \Theta}-\hat{\bf \Theta}_i(D_i)$, and 
the 3D parameter vector includes the source luminosity distance, the cosine of its declination and its right ascension, i.e.~${\bf \Theta}=(d_L,\cos\theta_{gw}, \phi_{gw})$.
The vector ${\bf \hat{\Theta}}_i(D_i)=(\hat{d}_L,\cos\hat{\theta}_{gw}, \hat{\phi}_{gw})_i$ defines the best measured values of the parameters (i.e.~the center of the Gaussian), which depend on the observed data. We assume that these are the only quantities measured from the data and drop the explicit dependence on $D_i$ in subsequent equations. The measurement uncertainties and correlations are described by the 3D correlation matrix ${\bf \Sigma}$, which we extract from the full fifteen-dimensional correlation matrix of the EMRI parameter estimation carried out in~\citet{Babak:2017tow}, by marginalising over all other parameters.
In practice, these posterior widths could depend on both the true parameters of the EMRI and on the particular realisation of the noise in the LISA data. For the well localised EMRIs we select in our analysis, these dependencies are likely to be weak.
Therefore, we assume here that the measurement uncertainties are fixed and known for each event. The first task is to associate a true host to each EMRI, consistent with these errors. To pick the true host within the sky localisation error $\Delta \Omega$, we first leave $(\cos\theta_{gw}, \phi_{gw})$ free, we compute $d_L$ for all galaxies in the Millennium sky and then evaluate the marginalised luminosity distance likelihood given by:
\begin{equation}
  p(d_L|D_i)=\frac{\exp\left\{-\dfrac{1}{2} \dfrac{\bigl(d_L-{\hat{d}_{L}}(D_i)\bigr)^2}{\sigma_{d_L}^2+\sigma_{WL}^2}\right\}}{\sqrt{2\pi\bigl(\sigma^2_{d_L}+\sigma^2_{WL}\bigr)}}\text{,}
\label{eq:dl}
\end{equation}
at the luminosity distance of each galaxy. A host galaxy is then randomly selected from the catalogue with probability proportional to this marginalised likelihood. 
Leaving $\cos\theta_{gw}$ and $\phi_{gw}$ free ensures that the pool of galaxies from which we select the host is large enough that the selection process reflects the clustering properties of galaxies at the measured distance of the event, while the luminosity distance of the selected host, $d_{L,H}$, is fully consistent with the LISA measurement error distribution by construction~\footnote{This is necessary because the EMRI catalogues used in this work were generated independently from the Millennium sky and its clustering properties. By allowing hosts to be selected within the whole ${4/3}\pi[(d_L+2\sigma_{d_L})^3-(d_L-2\sigma_{d_L})^3]$, we ensure that at least the hosts probability reflects the clustering of the Millennium sky within that luminosity distance range.}.
The host is described by the parameter vector ${\bf \Theta}_H=(d_{L,H},\cos\theta_{gw,H}, \phi_{gw,H})$.
We then fix $d_L=d_{L,H}$ in Eq.~\eqref{eq:psky} and draw a pair $(\cos\theta'_{gw}, \phi'_{gw})$ from the 2D slice of the $p({\bf \Theta})$ distribution. The LISA error volume is then re-positioned so that $\cos\hat{\theta}_{gw}=\cos\theta_{gw,H}-\cos\theta'_{gw}$ and $\hat{\phi}_{gw}=\phi_{gw,H}-\phi'_{gw}$, which, together with $\hat{d}_L$, redefine the vector ${\bf \hat\Theta}$. 
This procedure ensures that the true host is drawn consistently with the clustering properties of galaxies, and that its position relative to the best LISA measured values ${\bf \hat\Theta}$ is drawn according to the 3D volume error described by Eq.~\eqref{eq:psky}.
Now that we have identified a true host and have placed the LISA error volume appropriately with respect to it, we can truncate the galaxy catalogue by sub-selecting all galaxies $g_j$ so that $z_{\rm obs,{\it j}}$ is consistent with the LISA measurement, including the uncertainties in the cosmological parameters. 
In principle, the sum over galaxies in Eq.~(\ref{eqn:single_event_likelihood}) includes all galaxies in the catalogue up to $z=1$.
However, galaxies lying in the tail of the Gaussian distribution for all choices of cosmological parameters in the prior are exponentially suppressed in the likelihood. Therefore, it is more efficient to remove these from the catalogue. We retain those galaxies for which $z^-<z_{\rm obs,{\it j}}<z^+$, where $z^-$ and $z^+$ are implicitly given by
\begin{equation}
  d_L\pm2\sigma_{d_L}=(1+z^\pm)d_0\int_0^{z^{\pm}} \! \! \frac{\textup{d}z'}{F^\pm(z')}.
  \label{eq:zminmax}
\end{equation}  
Here $d_0 = c/H_0$ is the Hubble distance, and $F(z)=hE(z)$, with $E(z)$ given in Eq.~\eqref{eqn:E_z}. $F^\pm(z)$ are the realizations of $F(z)$ that minimize and maximize the $d_L-z$ conversion within the assumed prior on the cosmological parameters.
In practice Eq.~\eqref{eq:zminmax} extends the $\Delta{z}$ due to the $2\sigma$ error in the LISA measurement of $d_L$ as much as allowed by the prior range on the cosmology. 
Finally, we need to keep in mind that due to peculiar velocities, the true cosmological redshift of each galaxy $z_{\rm cos} \neq z_{\rm obs}$, the difference between the two being $\Delta{z}_{v_p}$. We thus consider all galaxies with $z_{\rm obs}\in[z^{-}-\Delta{z^-_{v_p}},z^{+}+\Delta{z^+_{v_p}}]\equiv\Delta{z_{\rm tot}}$, where $\Delta{z^-_{v_p}},\Delta{z^+_{v_p}}$ are simply $\Delta{z}_{v_p}$ calculated at $z^+, z^-$, taking a characteristic value of $v_p=500$ km s$^{-1}$, consistent with the standard deviation of the galaxy radial peculiar velocity distribution in the Millennium run~\citep{2005Natur.435..629S}.

We compute the weights $w_j$, appearing in Eq.~\eqref{eqn:single_event_likelihood}, for each galaxy $g_j$ within $\Delta{z_{\rm tot}}$ by marginalization of Eq.~\eqref{eq:psky} over $d_L$ (assuming uniform priors): 
\begin{equation}
p(D_i |{\rm cos}\,\theta_{\it j}\, \phi_{\it j} )=\int \!{\rm d}d_L\,p(D_i |{\bf \Theta})\equiv w_j \,,
\label{eq:margweights}
\end{equation}
evaluated at $(\rm{cos}\,\theta_{\it j}, \phi_{\it j})$
(where $w_j$ must not be confused with the DE equation of state parameters $w_0$ and $w_a$).
We do so for all galaxies falling into a sky region $\Delta\Omega$ covering $2\sigma$   
in the determination of $\cos\theta_{gw}$ and $\phi_{gw}$. In practice we assign probabilities only based on the marginalized 2D sky location error, discarding further information included in the correlation of $\rm{cos}\,\theta_{\it j}$ and $\phi_j$ with $d_L$, which is a conservative assumption (since we are overestimating the width of the marginal likelihood).

In summary, to assess the power of EMRI cosmology with LISA, we take three EMRI population models (M1, M5, and M6) from~\citet{Babak:2017tow}, and we consider the two cosmology scenarios presented in Sec.~\ref{sec:gw_cosmology_without_em_counterparts}: $\Lambda$CDM and DE.
We select only EMRIs with SNR$>$100 and associate to each event candidate hosts according to the procedure outlined above. 
For each model, the procedure of drawing the true host is repeated three times for each EMRI, so that we have three independent realizations of the galaxy host candidates, making our analysis robust with respect to statistical fluctuations in the galaxy distribution. 
We identify the independent realizations with a number following the name of the model, so that for model M1 we refer to M1\_1, M1\_2, and M1\_3, the same being for M2 and M3.
Finally, for each model, we investigate the impact of the mission lifetime by considering observation of EMRIs carried out in 4 years and 10 years of LISA operations.

An example of the outcome of the procedure outlined above is shown in Fig.~\ref{fig:galaxy}, where we depict the properties of the selected galaxies within the error box $\Delta{z_{\rm tot}}\times\Delta{\Omega}$ for three EMRI events selected from model M1\_2 under the assumption of a $\Lambda$CDM cosmology.
The $\cos{\theta}-\phi$ correlation is visible in the left column panels, together with the clustering pattern of the galaxies in the sky. 
Note that the true host, which differently from~\citet{MacLeod:2007jd} we do not remove from the group of possible hosts and is marked by the red dot in the left panels and the red vertical line in the middle panels, is generally offset from the center of the error box by construction, which mimics a realistic situation. 
The various elements of the $\Delta{z_{\rm tot}}$ construction procedure can be appreciated in the central panels.
The dark-green line marks the redshift corresponding to the best fit to the LISA distance measurement $\hat{d}_L$. The light-green lines bracket the $\Delta{z}$ interval associated to the $2\sigma_{d_L}$ 
LISA error assuming the true (i.e., the Millennium run) cosmology.
Finally, the yellow vertical lines bracket the interval $\Delta{z_{\rm tot}}$, allowed by the prior range on the cosmological parameters and accounting for the errors due to peculiar velocities (which are generally subdominant, unless $z\ll0.1$). 
The black histograms show the probability distributions of observed redshifts $z_{\rm obs}$ of the hosts in the error volume, where the probability of each individual host, $w_j$, is given by Eq.~\eqref{eq:margweights}. As it should be, $z_{\rm obs}$ of the true host generally falls within the light-green lines, although this is not always strictly the case due to peculiar velocities. 
Finally, the right panels show sanity checks of the offset distributions of $d_L$, $\cos\theta$, and $\phi$ of the candidate hosts from the best-measured LISA values normalized to the respective measurement errors. 
The distributions for $d_L$ (black histograms) extend to several $\sigma$ due to the extra contribution allowed by varying the cosmological parameters within the prior range.

The top panels in Fig.~\ref{fig:galaxy} show a ``very good EMRI'', for which the putative hosts cluster in the $\Delta{z}$ range allowed by the true cosmology. 
The middle panels show a ``non informative EMRI'', displaying an essentially flat (although very noisy) probability distribution of the hosts across the whole redshift range allowed by the cosmological prior. 
Finally, the bottom panels show a ``bad EMRI''. In this latter case, the true host was picked in a small group of galaxies at $z\approx 0.35$, and by chance, in the same sky region, there is a much larger group of galaxies at $z\approx 0.45$, which skews the probability distribution of the hosts towards a cosmology different from the true, although still within the allowed prior. Events of this type are expected to be subdominant, and by stacking the posteriors of several events, the true cosmological parameters naturally emerge from the analysis, as we now show.


\section{Results} 
\label{sec:results}

In this section we report the results of our investigation following the procedure described in the previous sections.
We will first show the results for the $\Lambda$CDM scenario (Sec.~\ref{sub:lcdm}) and then for the DE scenario (Sec.~\ref{sub:dark_energy}).
For both scenarios we will consider 4 and 10 years of LISA observational time.

Now we describe how we obtain the posteriors reported in this section. Concerning the 4-year results, in order to accumulate enough statistical evidence to produce reliable results for a 4-year LISA mission, for each EMRI model we generated three 4-year catalogs from each of the three realisations of the 10-year catalog described in Sec.~\ref{sec:EMRIsel} and~\ref{sec:cross_correlation_with_galaxy_catalogues}. 
The number of events in each 4-year catalog was obtained as follows: i) we randomly picked an integer $n_{4\mathrm{yr}}$ by drawing from a Poisson distribution with mean equal to the 4/10 of the total number of events (see Table~\ref{tab:Nemri}); ii) we then selected randomly $n_{4\mathrm{yr}}$ events from the original catalog; iii) finally we applied the SNR$>$100 cut to obtain the final set of EMRIs for the analysis.
In this way we have a total of nine 4-year realisations per EMRI scenario (for example, in case of M1 we produced three realisations for M1\_1, three for M1\_2, and three for M1\_3), each of which has been analysed separately. 
The posteriors for the cosmological parameters are then averaged~\citep{DelPozzo:2017kme} over the different realisations. This procedure is applied for both the $\Lambda$CDM and the DE cosmological inference models. 
Our results with 10 years of LISA mission lifetime for each EMRI scenario are simply obtained averaging the analysis of the three different catalog realisations (again, for M1 it will be the average of M1\_1, M1\_2, and M1\_3). 
Such catalogs contain the exact number of EMRI events as given in the fifth and sixth columns of Table~\ref{tab:Nemri}, depending on the assumed cosmological model.

\begin{figure*}
\includegraphics[width=0.95\textwidth]{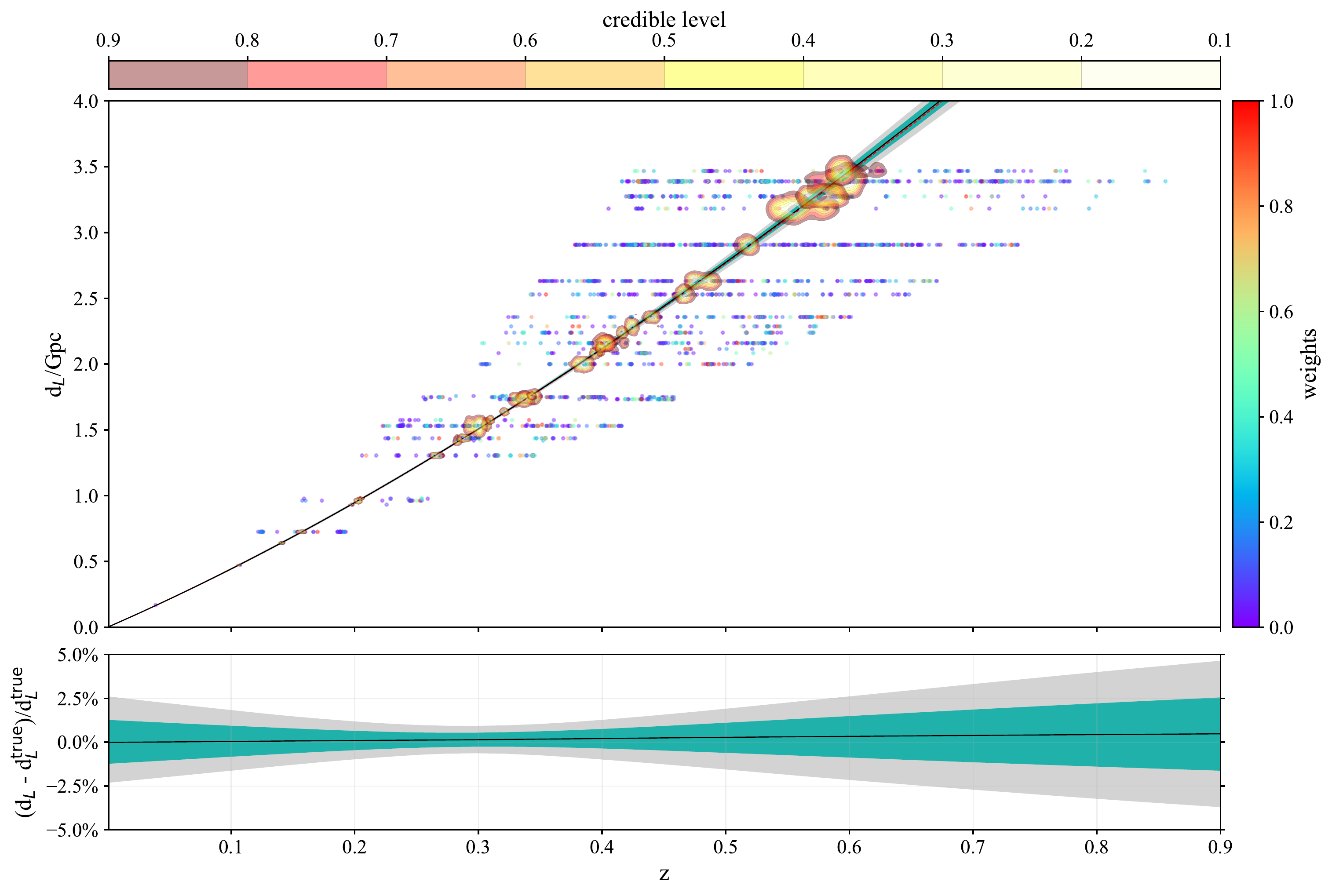}
\caption{
$d_L-z$ regression line for M1\_2, one of the three realisations of our fiducial scenario M1 in the 10-year mission case: median (solid black) and 68\% and 95\% credible regions in light seagreen and light gray, respectively. The dashed line corresponds to the Millennium Simulation fiducial values of $h$ and $\Omega_m$. The coloured regions show the posterior distribution for $z_{gw}$ and $d_L$ for each EMRI event. The horizontal dots show the redshift of each candidate galaxy host for that particular EMRI. For illustrative purpose, we assigned to each galaxy a luminosity distance equal to $\hat{d}_L$. The dots are also colour-coded from violet to red for increasing values of the weights $w_j$. The bottom plot shows the residuals of the inferred regression line credible regions, illustrating the accuracy in $d_L$ as a function of the redshift for M1\_2.}
\label{fig:regression}
\end{figure*}

For illustrative purposes in Fig.~\ref{fig:regression} we show a distance-redshift diagram as obtained from our analysis in the $\Lambda$CDM scenario for the M1\_2 realisation only of our fiducial EMRI model, assuming 10 years of LISA mission lifetime.
The credible regions pictured in Fig.~\ref{fig:regression} for each EMRI event are the posterior distributions resulting from combining the LISA distance measurement with the combined information collected from all galaxies within the sky localisation region associated to that particular event.
For each EMRI, to compare the prior range for $z_{gw}$ with the corresponding range of galaxy host redshifts, in Fig.~\ref{fig:regression} we show the redshift $z_j$ of each galaxy $j$ within the sky localisation region of each EMRI event, displayed  as $w_j$-weighted dots at a fixed luminosity distance set, for convenience, equal to $\hat{d}_L$.  
Note also how measurements at low redshift are more accurate due to the lower number of galaxies contained within the reconstructed volume, while events at higher redshift possess on average less constraining power due to the large number of galaxies within their sky error boxes.

\subsection{$\Lambda$CDM} 
\label{sub:lcdm}

\begin{figure*}
    \centering
    \begin{tabular}{ccc}
        \vspace{0.2cm}
           & \textbf{$\Lambda$CDM, 4yr} &   \\
         Model M5 {\it (pessimistic)} & Model M1 {\it (fiducial)} & Model M6 {\it (optimistic)} \\
        \includegraphics[width=0.32\textwidth]{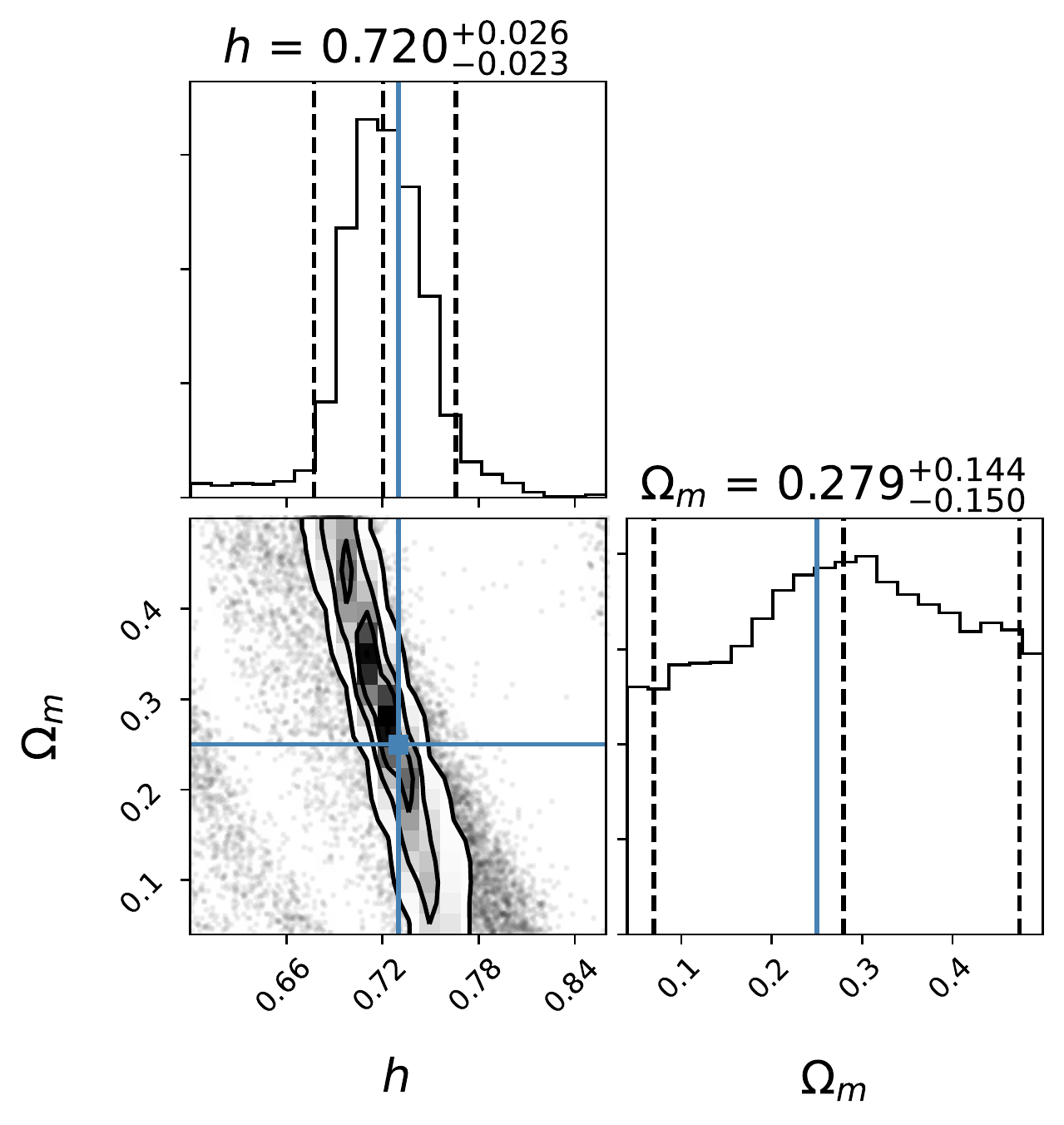}&
        \includegraphics[width=0.32\textwidth]{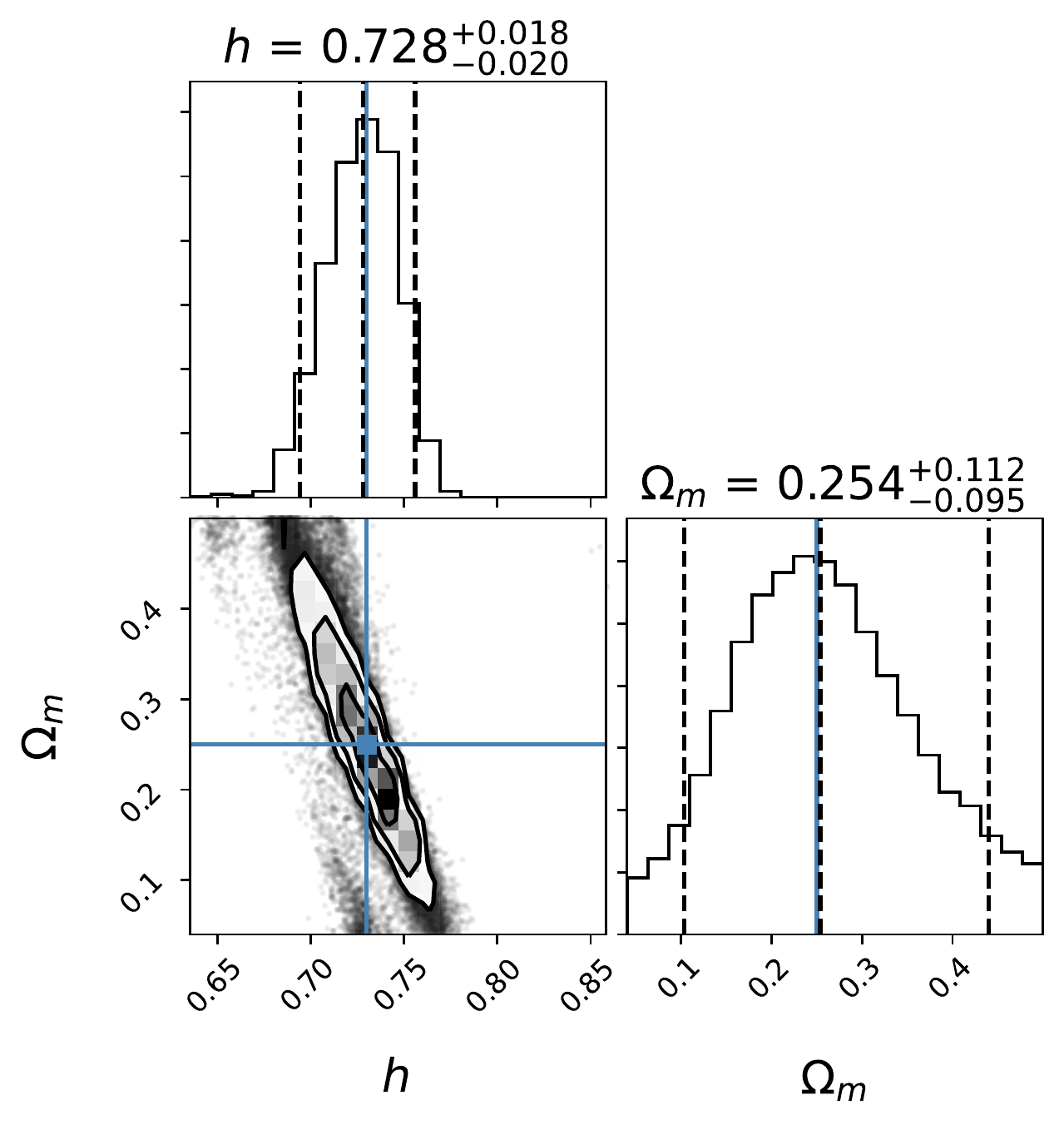}&
        \includegraphics[width=0.32\textwidth]{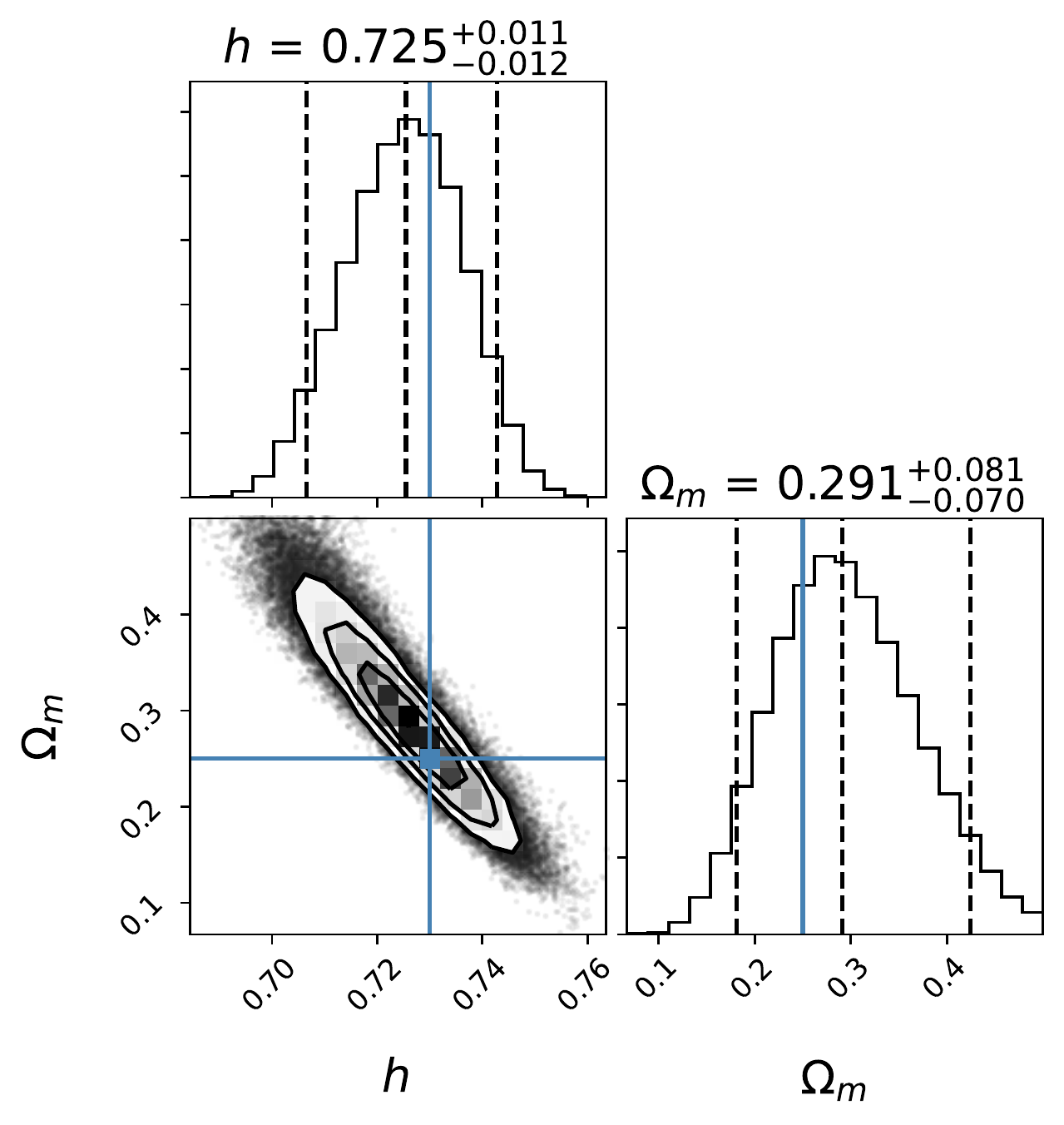}\\
        \vspace{0.2cm}
           & \textbf{$\Lambda$CDM, 10yr} &   \\
         Model M5 {\it (pessimistic)} & Model M1 {\it (fiducial)} & Model M6 {\it (optimistic)} \\
        \includegraphics[width=0.32\textwidth]{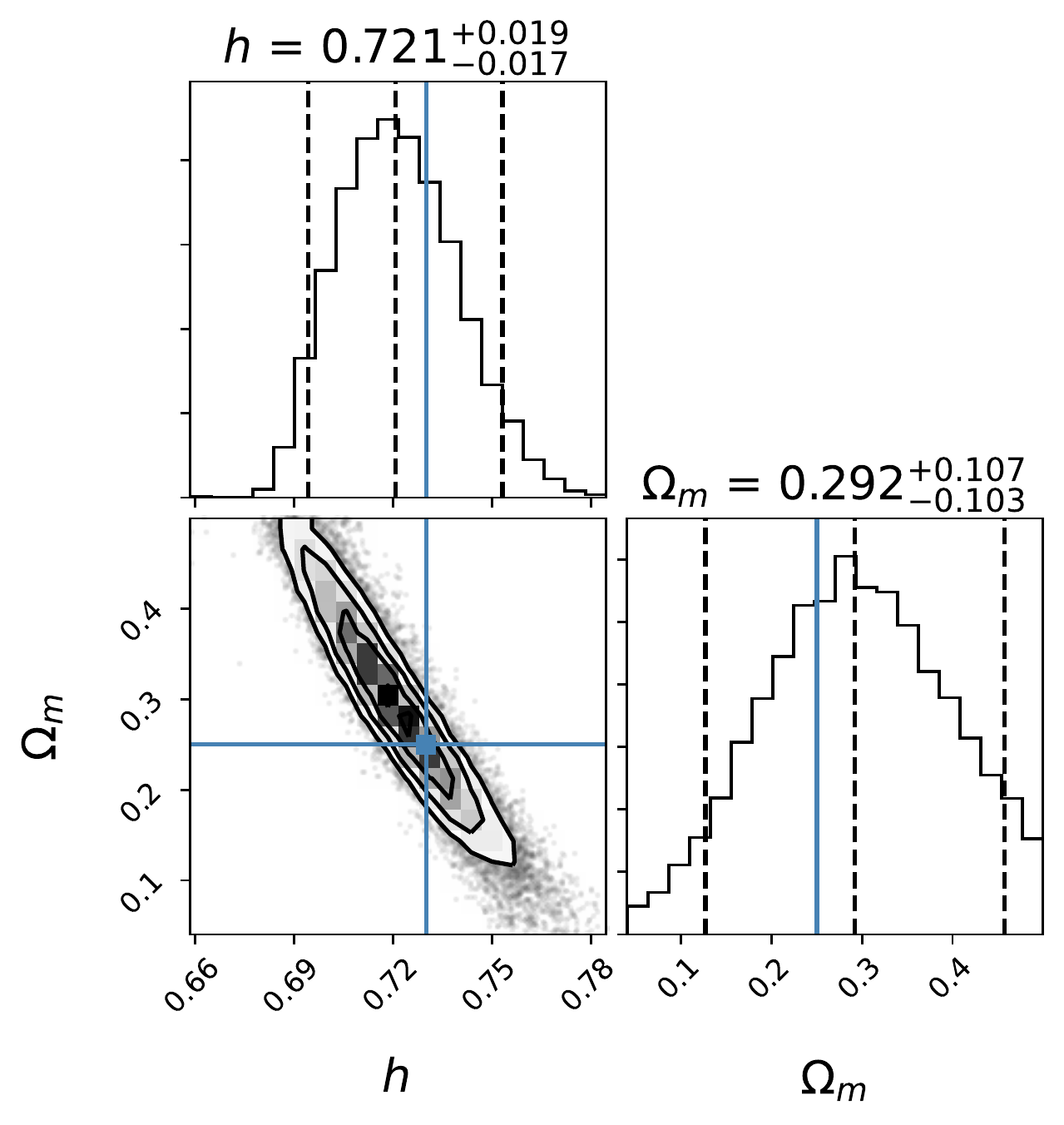}&
        \includegraphics[width=0.32\textwidth]{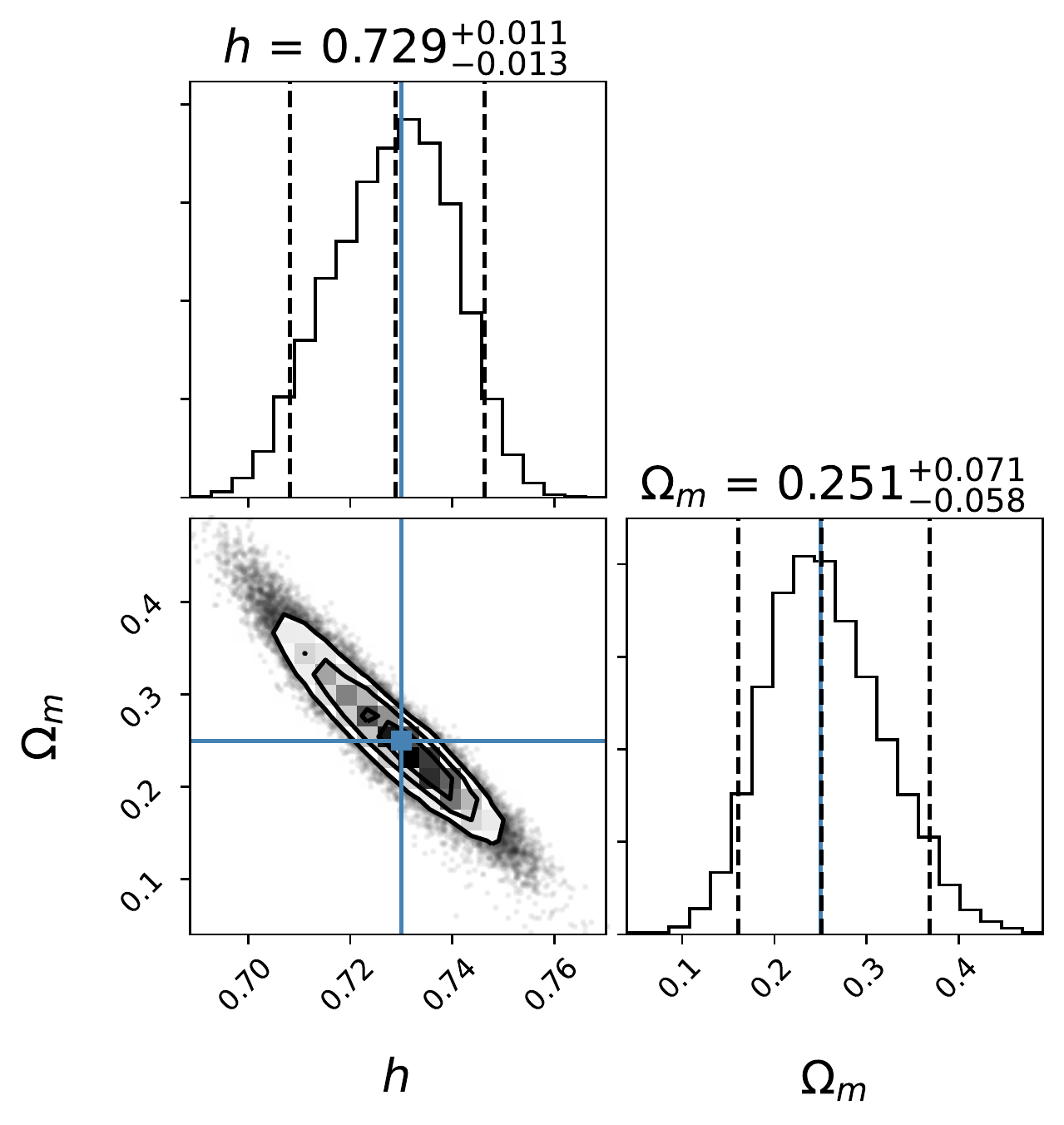}&
        \includegraphics[width=0.32\textwidth]{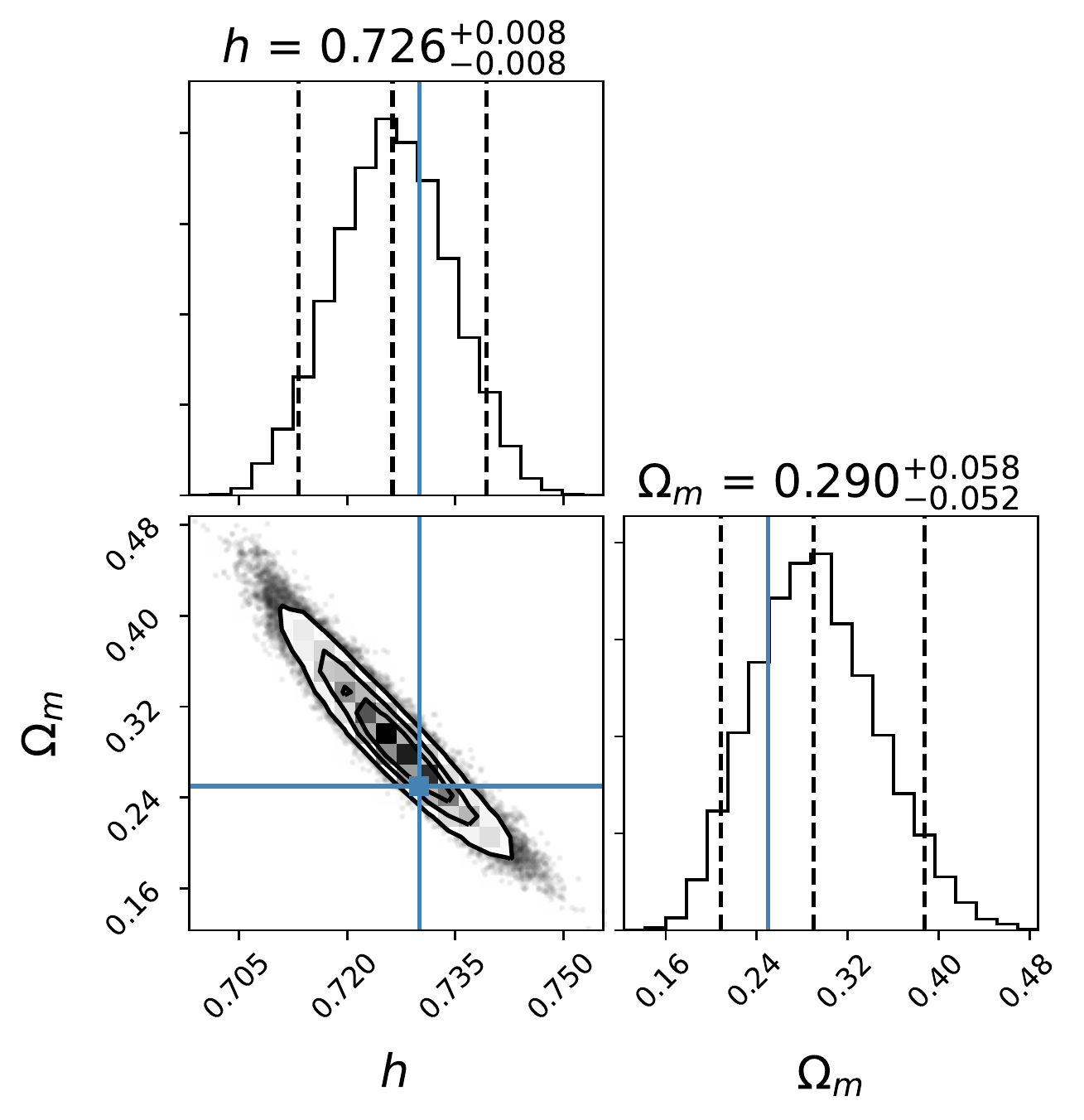}\\
    \end{tabular}
    \caption{
    Corner plots of the posteriors for the parameters $h$ and $\Omega_m$ in the $\Lambda$CDM scenario (4 and 10 years of observations). In each plot the lower-left panel reports the contours of the joint posterior, while the upper and right panels show the same posterior after marginalization over one parameter. In each panel, the cyan lines mark the fiducial values and the black dashed lines indicate the median and 90\% credible interval extracted from the marginalized posterior.
    }
    \label{fig:corner_lcdm}
\end{figure*}

Results for the $\Lambda$CDM scenario are reported in Fig.~\ref{fig:corner_lcdm} for our three selected EMRI models: M5 (pessimistic), M1 (fiducial), and M6 (optimistic).
We show corner-plots with 2D posteriors in the $h$-$\Omega_m$ parameter space, together with marginalized 1D posteriors on both $h$ and $\Omega_m$.
Results with both 4 years (upper-row) and 10 years (lower-row) of LISA mission operation are shown.
The constraints reported in the figure indicate 90\% confidence levels around the measured median value.
In what follows we will comment in detail on these two scenarios.

\subsubsection{4 years of LISA observations}

As described above, the joint posteriors reported in Fig.~\ref{fig:corner_lcdm} are averaged over the nine independent realizations of each EMRI scenario.

As reported in Fig.~\ref{fig:corner_lcdm}, the constraints we obtain on $h$ with 4 years of LISA observations range  from a maximum of 0.026 (3.6\%), within the M5 scenario, to a minimum of 0.012 (1.6\%), within the M6 scenario.
The constraints on $\Omega_m$ go instead from 0.150 (60.0\%) in the M5 scenario, to 0.081 (32.0\%) in the M6 scenario.
In general, the predominance of low-redshift events in the distance-redshift diagram is such that they are more efficient at constraining $H_0$ rather than $\Omega_m$.
Note that the posterior on $\Omega_m$ in the M5 (pessimistic) case is actually dominated by our prior, implying that in this scenario the LISA data do not yield statistically relevant information on $\Omega_m$.
Also note that, as expected, our results obtained with M1 (our fiducial scenario) are superseded by those obtained with M6 (optimistic), as in this case the total number of useful EMRIs is more than doubled (cf.~Table~\ref{tab:Nemri}).

\subsubsection{10 years of LISA observations}

As shown by the lower-row plots in Fig.~\ref{fig:corner_lcdm}, the constraints on $h$ now range from a worst-case result of 0.019 (2.6\%) to a best-case result of 0.008 (1.1\%), while the constraints on $\Omega_m$ range from 0.107 (42.8\%) to 0.058 (23.2\%).
Differently from the 4-year case, the posterior on $\Omega_m$ in the M5 scenario (pessimistic) is not dominated by the prior, meaning that in this scenario 10 years of mission duration could start yielding meaningful information on $\Omega_m$.

The results obtained with M5, our pessimistic scenario, might seem too optimistic if one considers the fact that only 6 EMRI events are counted in the 10-year catalog that we used (see Table~\ref{tab:Nemri}). However, we have selected the loudest events, which should be the most informative from a parameter estimation point of view.
An analogous consideration could be made for the 4-year results, where the number of EMRI events in the employed M5 catalogs are on average only $n_{4\mathrm{yr}}=2.4$.

We have moreover checked that the 10-year M5 catalog contains a low-redshift event which in the three versions of the 10-year catalog happens to have only 1, 3, and 8 galaxies within its 3D localisation error box.
In the catalog where this EMRI has a single host, the inference of its redshift is evidently optimal. In the catalog where it has three galaxies, two of these galaxies are spot-on the true redshift value, while the third one is less relevant, because it lies at the margins of the sky localisation region and thus, according to how our analysis is implemented (cf.~Sec.~\ref{sec:gw_cosmology_without_em_counterparts}), it is weighted with a lower probability with respect to the others.
This event alone therefore provides a highly accurate measurement of the Hubble constant at low redshift, which dominates the results reported in the first column of Fig.~\ref{fig:corner_lcdm}.
We note that although the detection of such an advantageous event is a question of luck, the actual probability of finding a similar event in the few EMRI events included in our catalogs is actually non-negligible, in that our analysis automatically selects for well-localized events at high SNR.
In general, any EMRI event of this kind will substantially drive our results, while less well-localized, low-SNR events are not expected to contribute (but see App.~\ref{sec:app_low_snr} for a test-case in which we \emph{only} use the faintest events of our fiducial model).

\subsection{Dark Energy} 
\label{sub:dark_energy}

\begin{figure*}
    \centering
    \begin{tabular}{ccc}
        \vspace{0.2cm}
           & \textbf{DE, 4yr} &   \\
         Model M5 {\it (pessimistic)} & Model M1 {\it (fiducial)} & Model M6 {\it (optimistic)} \\
        \includegraphics[width=0.32\textwidth]{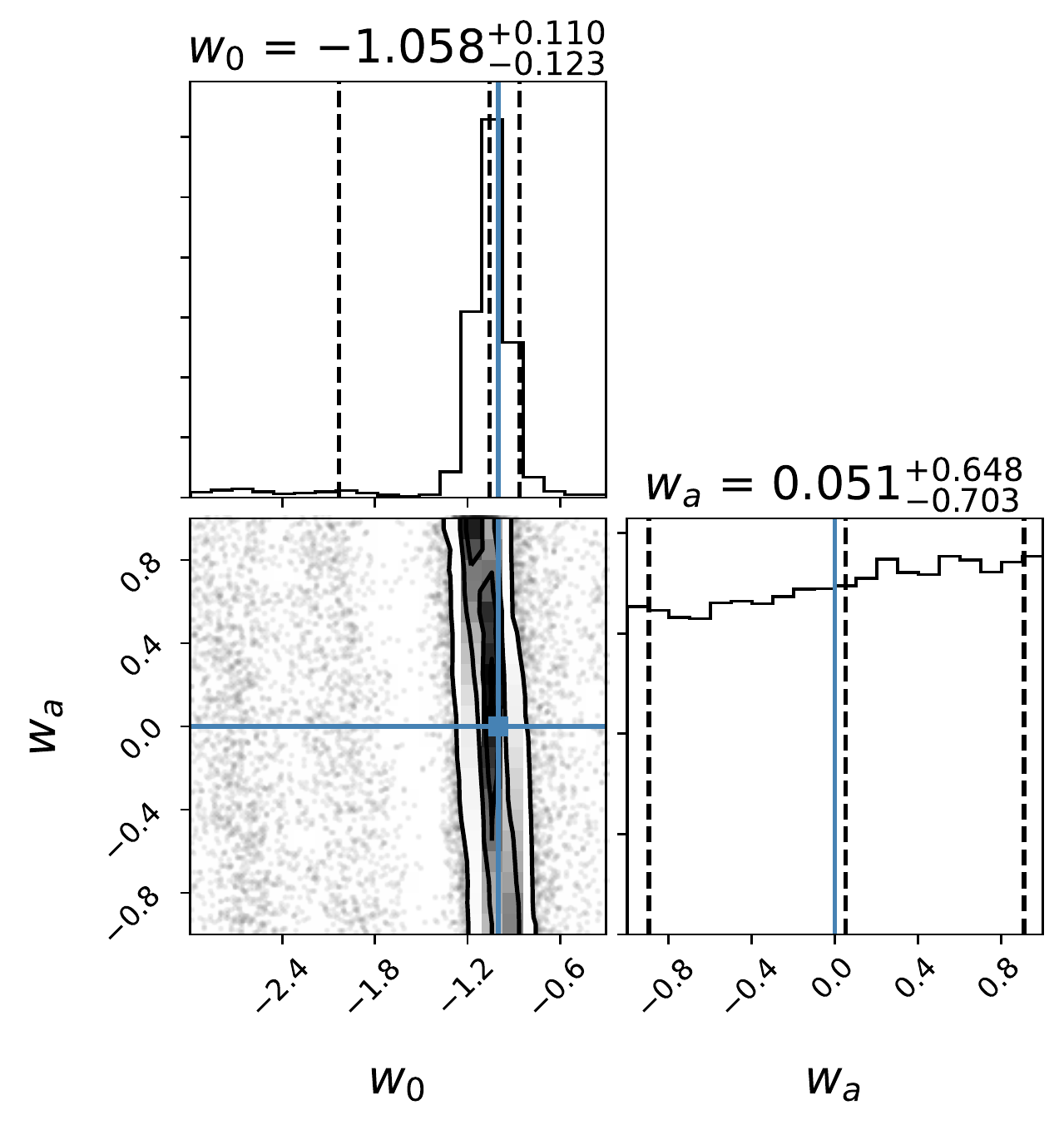}&
        \includegraphics[width=0.32\textwidth]{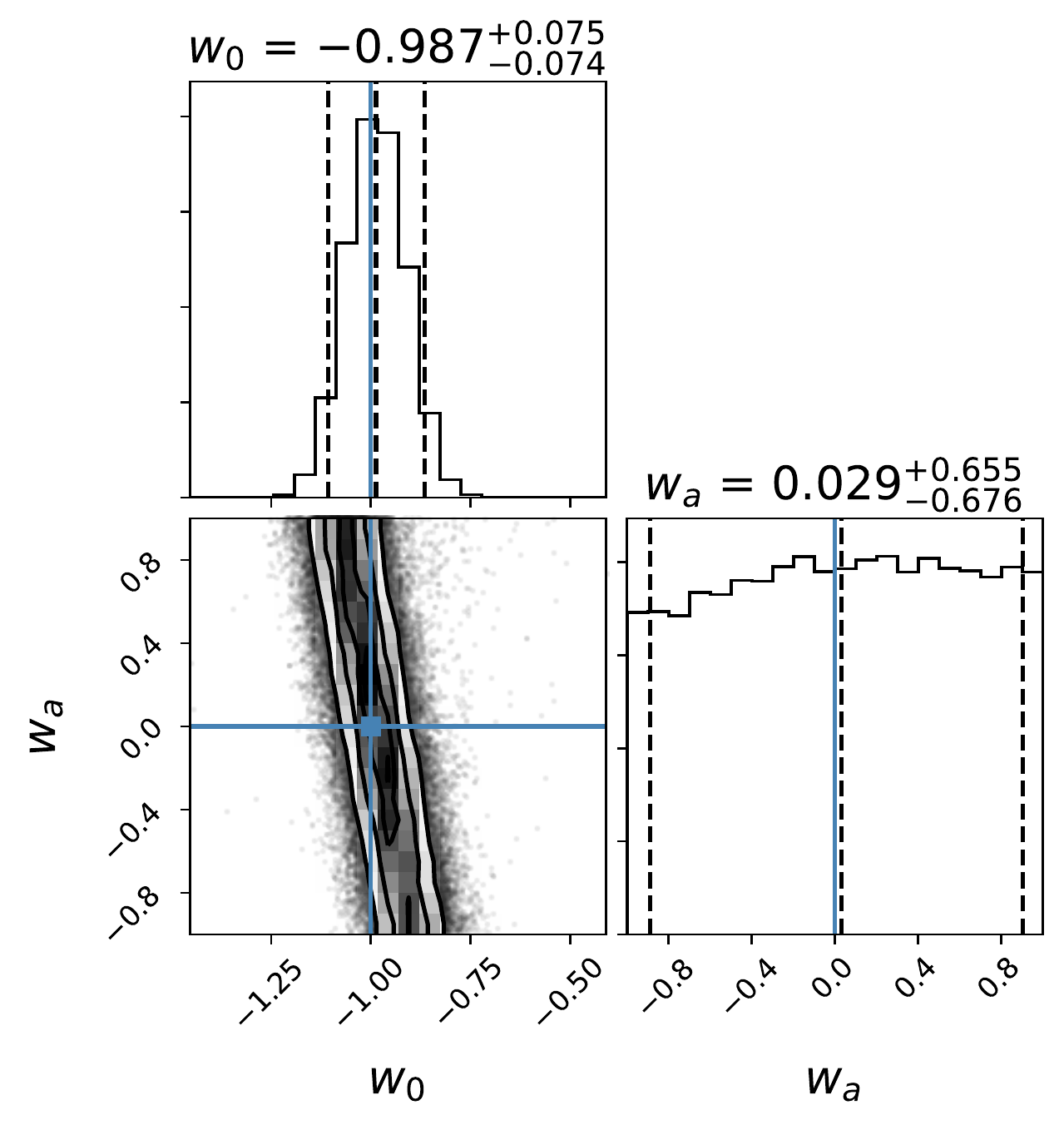}&
        \includegraphics[width=0.32\textwidth]{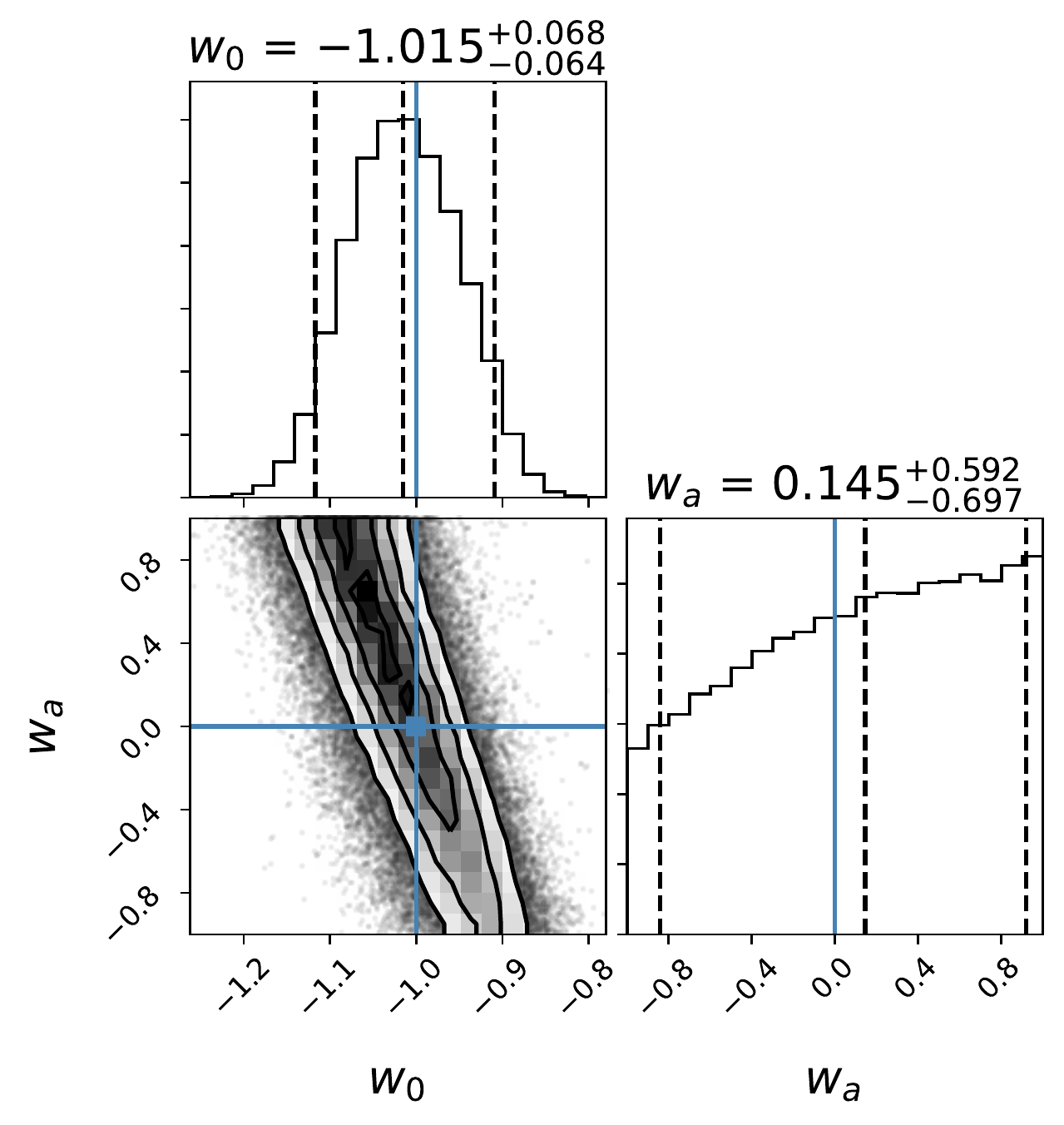}\\
        \vspace{0.2cm}
           & \textbf{DE, 10yr} &   \\
         Model M5 {\it (pessimistic)} & Model M1 {\it (fiducial)} & Model M6 {\it (optimistic)} \\
        \includegraphics[width=0.32\textwidth]{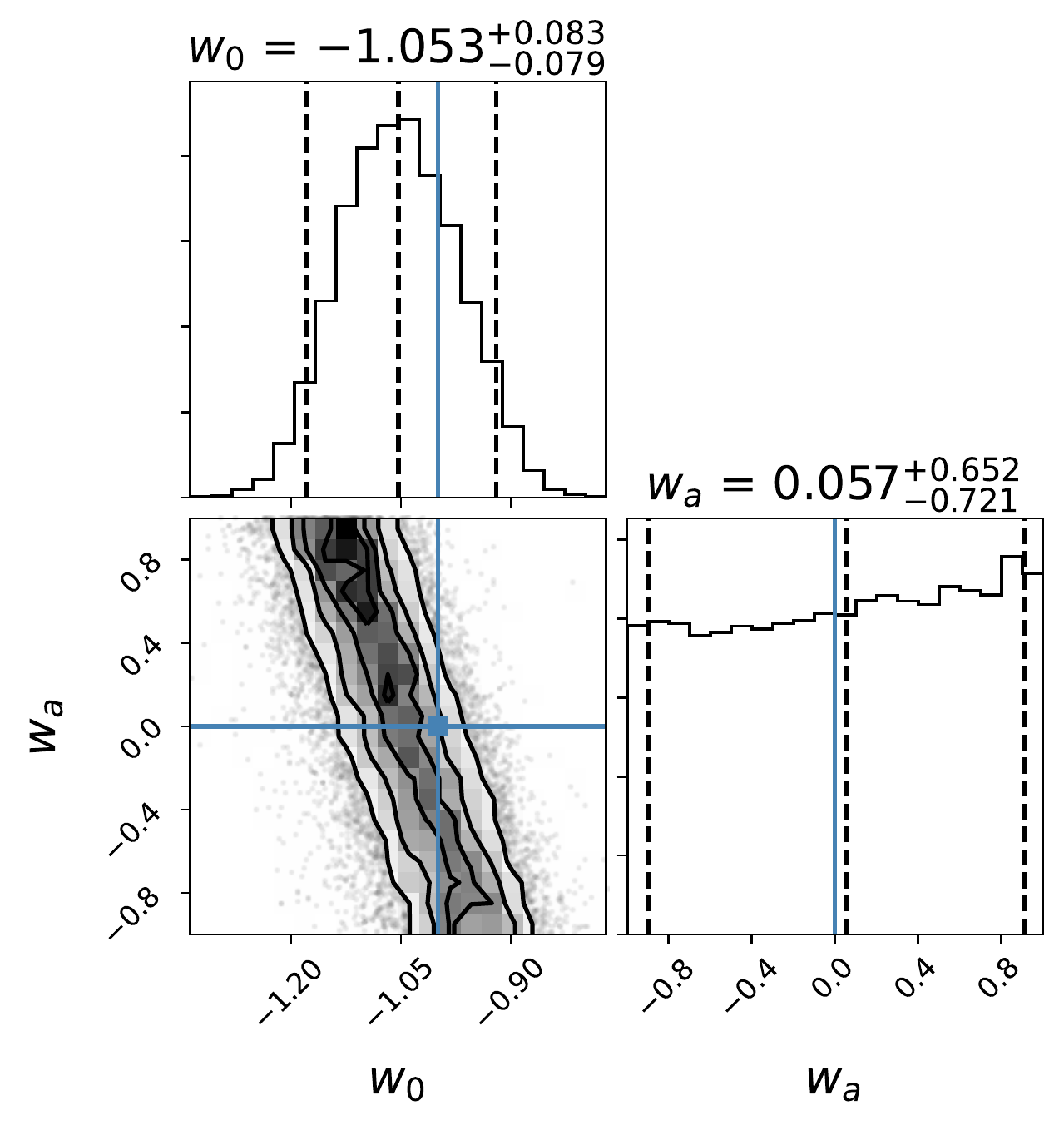}&
        \includegraphics[width=0.32\textwidth]{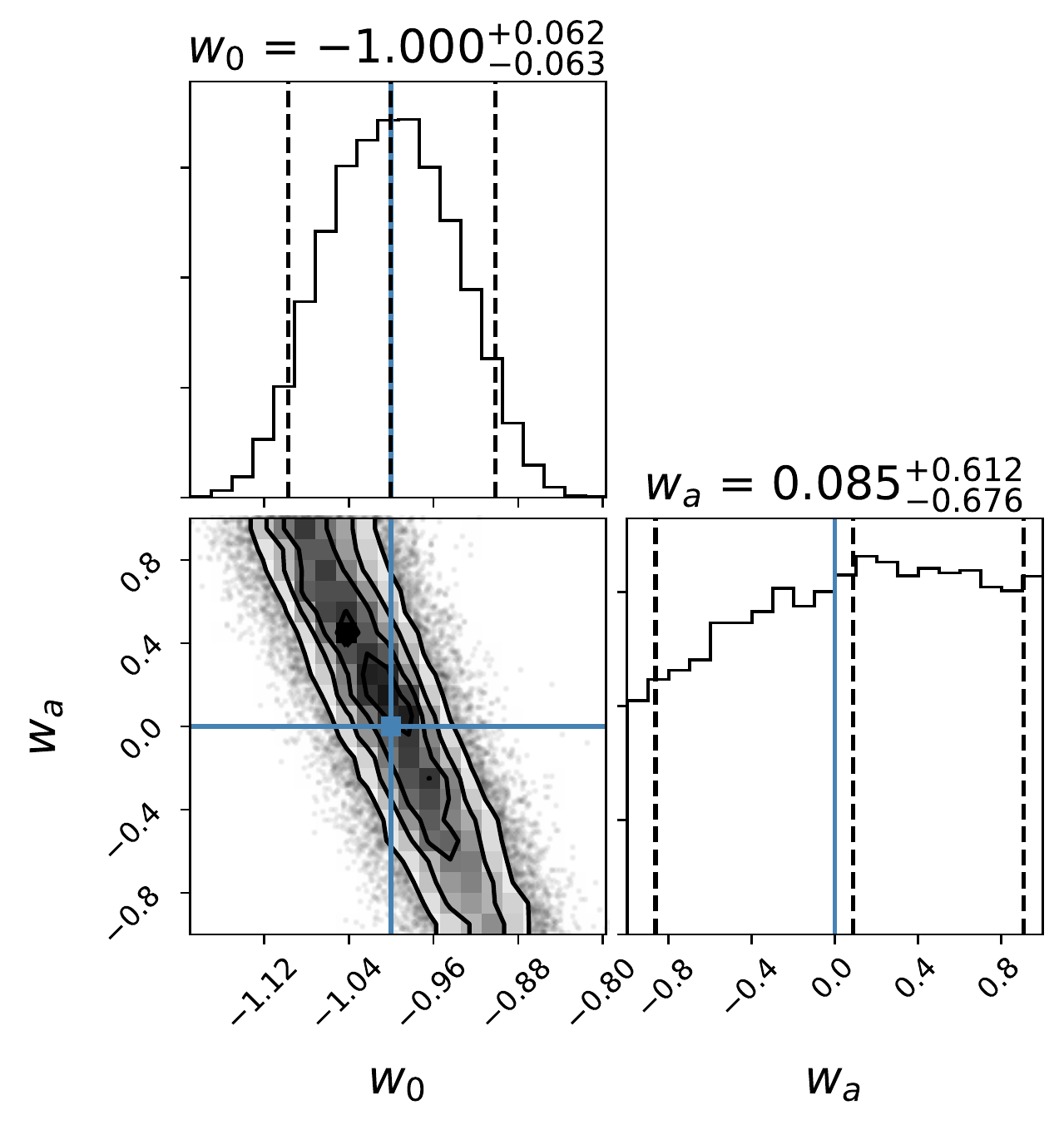}&
        \includegraphics[width=0.32\textwidth]{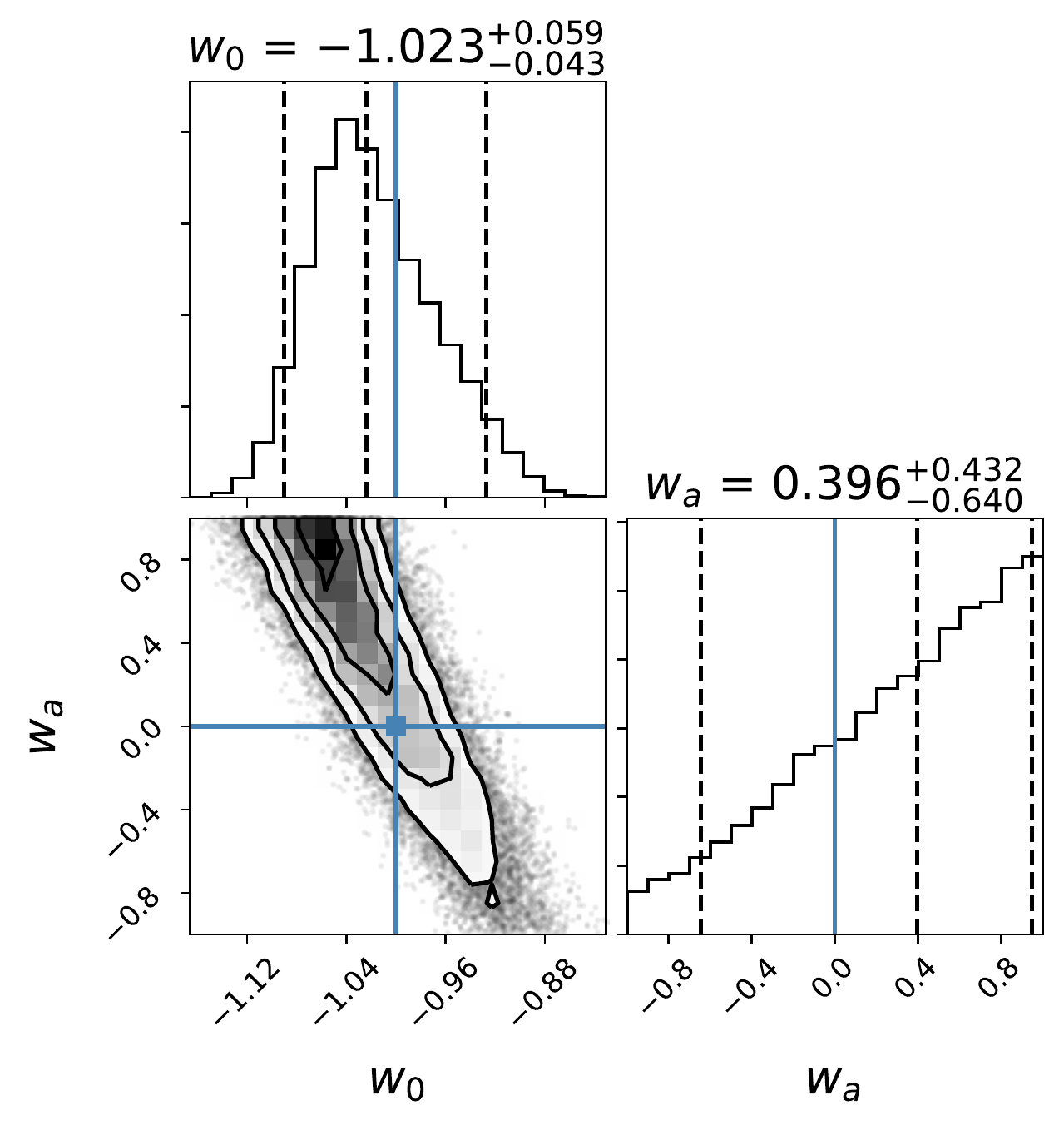}\\
    \end{tabular}
    \caption{Corner plots of the posteriors for the parameters $w_0$ and $w_a$ in the DE scenario (4 and 10 years of observations). Same as Fig.~\ref{fig:corner_lcdm}. See App.~\ref{sec:app_railing_wa} for a discussion on the M6 model.
    \label{fig:corner_de}
    }
\end{figure*}

Results for the DE cosmological scenario are reported in Fig.~\ref{fig:corner_de}, again for all three EMRI models.
We show corner-plots with 2D posteriors in the $w_0$-$w_a$ parameter space, together with marginalized 1D posteriors on both $w_0$ and $w_a$, for both 4 years (upper-row) and 10 years (lower-row) of LISA mission operations.
Again constraints shown in Fig.~\ref{fig:corner_de} define 90\% confidence levels around median values.
In what follows we will address in more details the two observational scenarios.

As already noted in Sec.~\ref{sec:cross_correlation_with_galaxy_catalogues}, the total number of useful events for each EMRI model slightly varies according to the cosmological model under consideration (cf.~Table~\ref{tab:Nemri}); 
thus the DE catalogs have a slightly different total number of useful EMRIs with respect to the ones used for $\Lambda$CDM.

\subsubsection{4 years of LISA observations}

From the top-row panels of Fig.~\ref{fig:corner_de} we can notice that constraints on $w_0$ range from a maximum of 0.123 (12.3\%) to a minimum of 0.068 (6.8\%) for the M5 (pessimistic) and M6 (optimistic) models, respectively.
On the other hand, the measurements on $w_a$ are largely uninformative as they yield 90\% confidence intervals of $\sim0.7$ (corresponding to $\sim70\%$ of the prior) irrespective of the EMRI cosmological scenario.
This implies that although LISA EMRIs might be able to tell us something interesting on the current value of the DE equation of state ($w_0$), they are likely to not have the statistical power to tell us anything about its current time evolution ($w_a$).

\subsubsection{10 years of LISA observations}

Results are reported in the bottom row of Fig.~\ref{fig:corner_de}.
We can see that with 10 years of observations LISA EMRIs will yield constraints on $w_0$ ranging from 0.083 (8.3\%) to 0.059 (5.9\%), slightly improving over the 4-year results.
Results for $w_a$ are again largely uninformative, as they still produce 90\% confidence constraints around 0.6-0.7, which still constitute a non-negligible fraction of the prior. However, we notice the apparent railing of $w_a$ in case of M6, which becomes prominent in the 10-year case. We will comment on this in App.~\ref{sec:app_railing_wa}.


\section{Discussion} 
\label{sec:discussion}

\begin{figure}
    \centering
    \textbf{Summary plots}\\
    \begin{tabular}{c}
        \vspace{0.2cm}
            \\
         $h$ \\
        \includegraphics[width=0.47\textwidth]{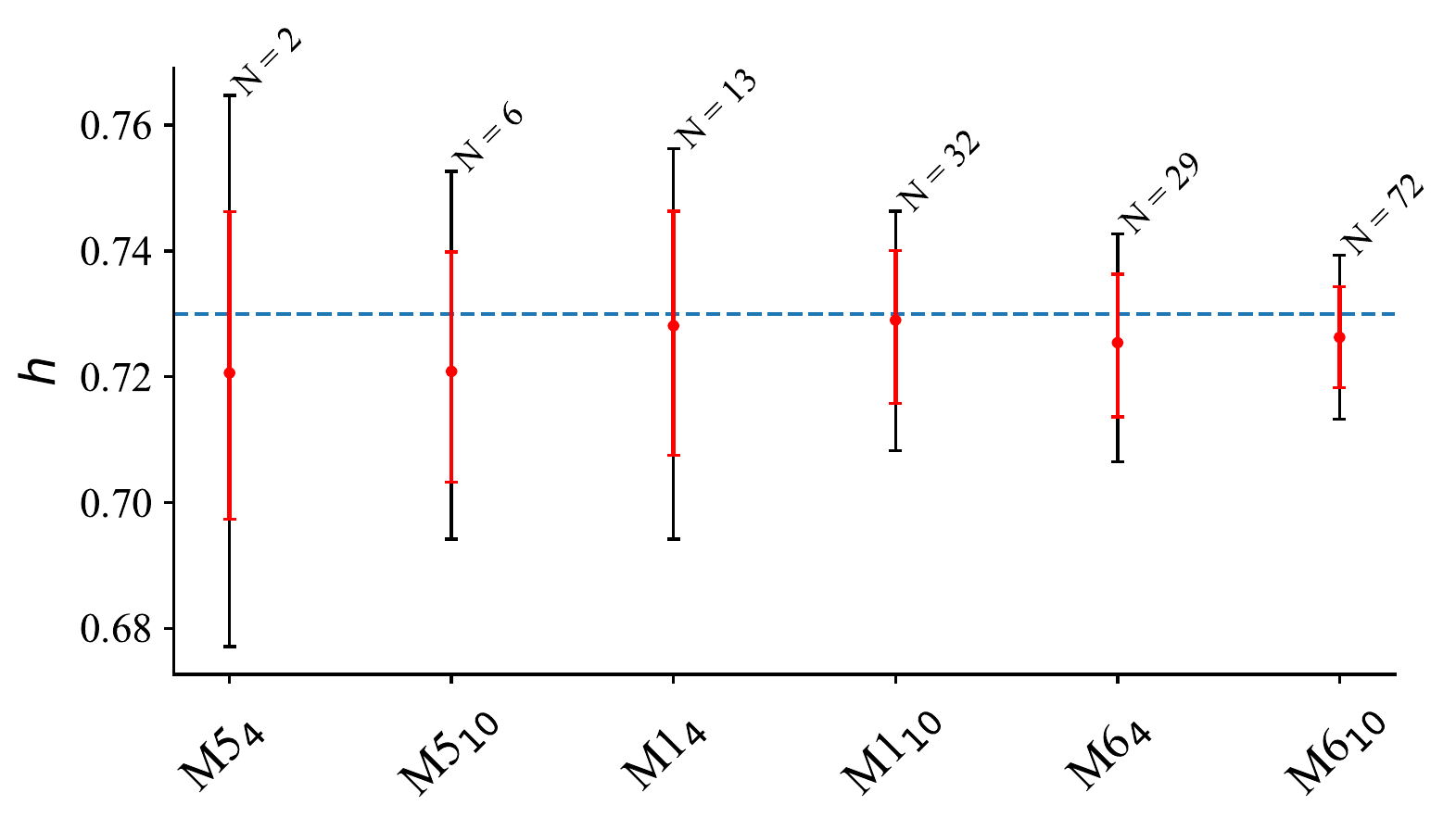} 
        \\
        \vspace{0.2cm}
            \\
       $\Omega_m$\\
        \includegraphics[width=0.49\textwidth]{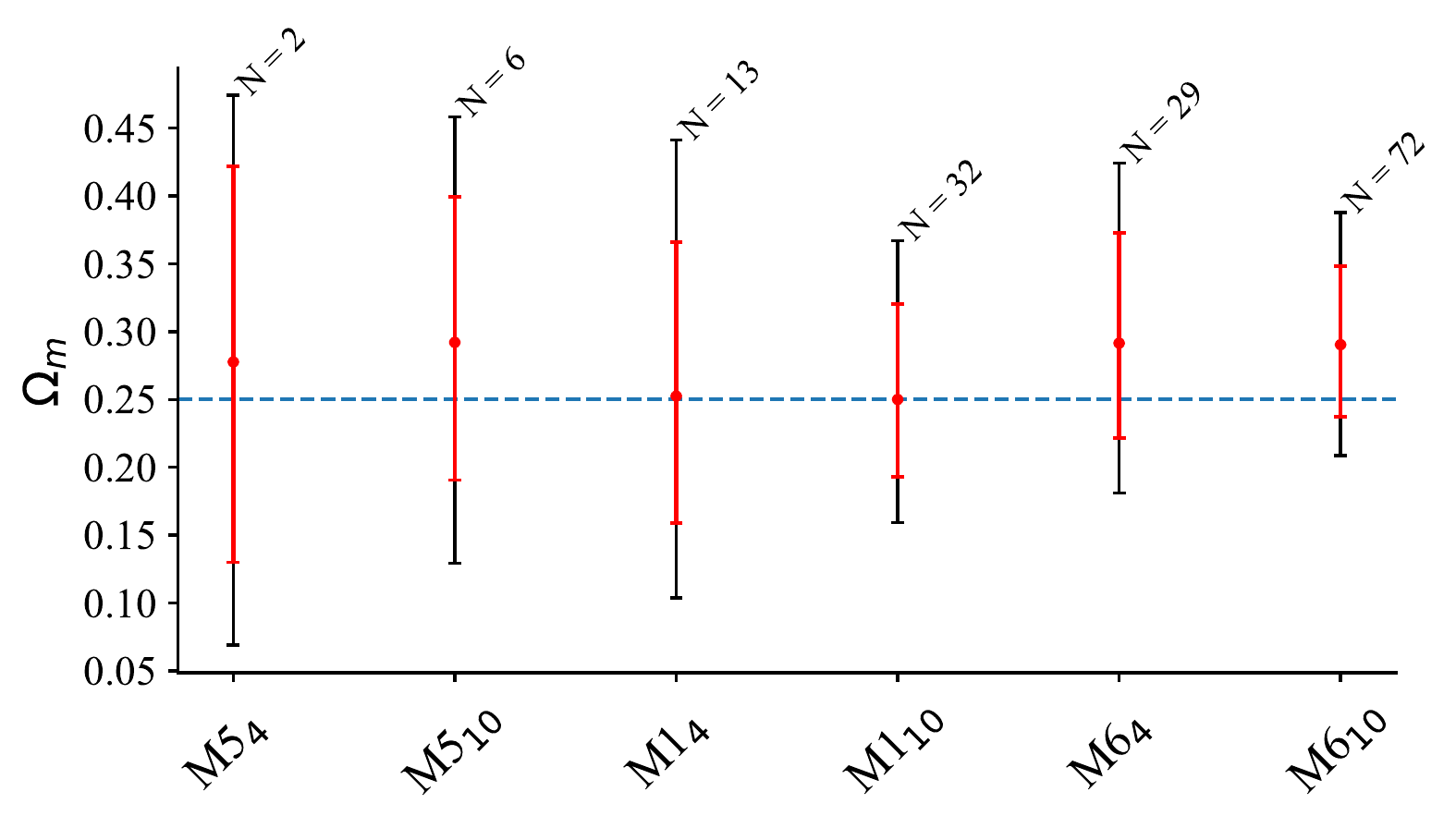}\\
        \\
                \vspace{0.2cm}
         $w_0$\\
        \includegraphics[width=0.49\textwidth]{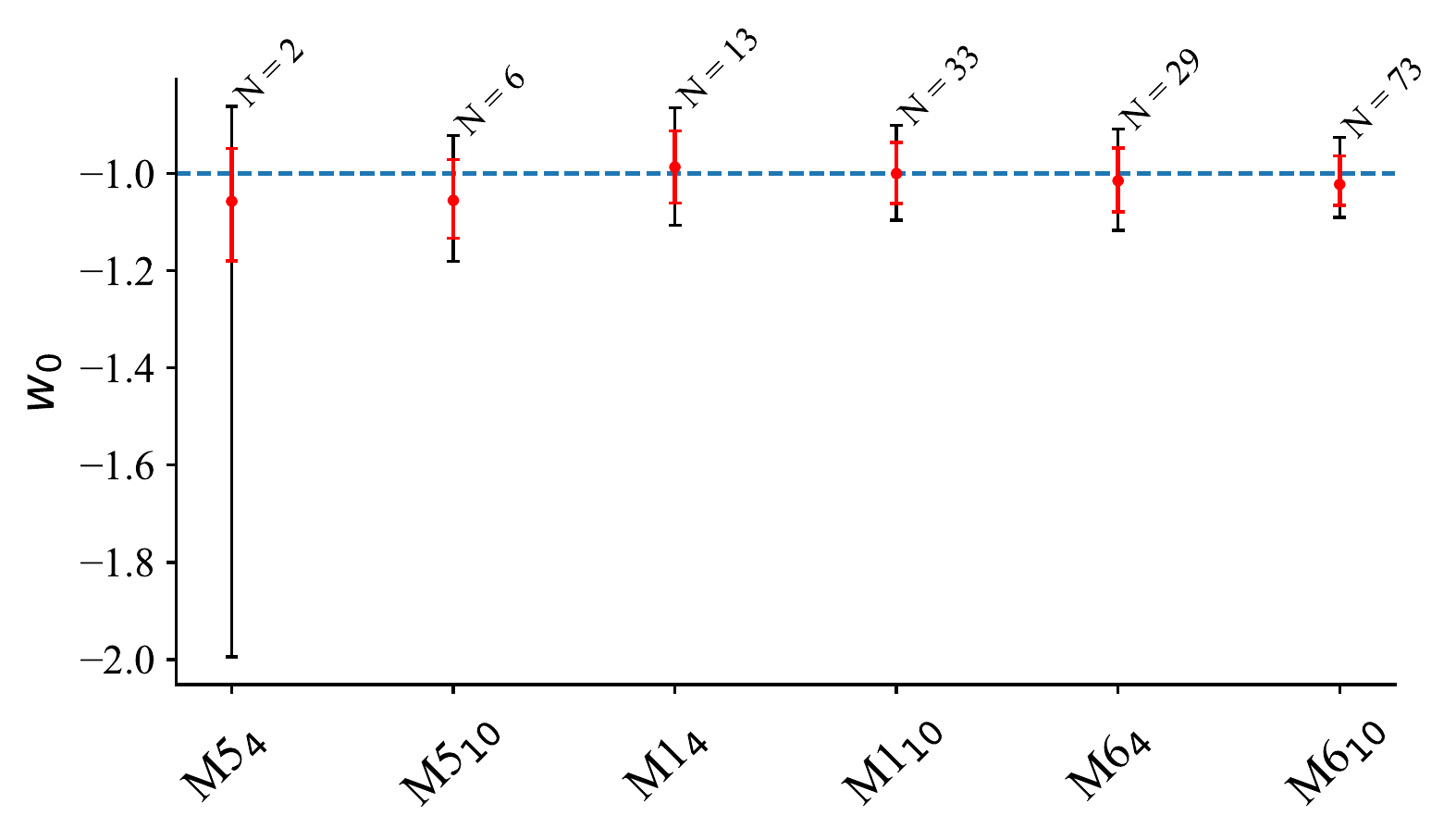}\\
    \end{tabular}
    \caption{
    Summary of results. We show 90\% (black) and 68\% (red) percentiles, together with the median (red dot) for both $h$, $\Omega_m$, and $w_0$ for each EMRI model (M5, M1, M6) and LISA observational scenario (4 and 10 years) considered. The blue dashed horizontal line denotes the true cosmology. For each  data point, we also report the average number $N$ of EMRIs considered in the analysis, see Table.~\ref{tab:Nemri}.
    }
    \label{fig:summary}
\end{figure}

Overall the results we obtained from our investigation show that EMRIs detected by LISA will have an interesting cosmological potential.
Assuming the $\Lambda$CDM model, the Hubble constant will be probed at the percent level, while in the evolving DE scenario the equation of state of DE will be tested with an accuracy better than 10\%. Some constraint, even though not particularly strong ($\sim20\%$), can also be put on $\Omega_m$.
The constraints on these three parameters, for all three EMRI models and both 4 and 10 years of LISA observations, have been summarised in Fig.~\ref{fig:summary}.
On the other hand, only a marginal gain will be achieved on $w_a$ (for DE), which usually is better constrained by measurements at high redshift, which are not so numerous in our catalogs.

\subsection{Comparison with EM observations}

How do our results compare with current and future EM observations?
Current EM measurements of the Hubble constant have reached percent levels of accuracy with observations collected from the CMB~\citep{Ade:2015xua,Aghanim:2018eyx} and from local-Universe distance indicators~\citep{Riess:2016jrr,Riess:2019cxk}, in particular type-Ia supernovae (SNIa).
However, as mentioned in the Introduction, these two techniques yield different values of the Hubble constant which are now in tension with each other at more than the 4$\sigma$ statistical level~\citep{Riess:2019cxk}.
Whether this tension is due to systematics in one of the two measurements or to new physics beyond $\Lambda$CDM, it will be only decided by future observations, including
not only new EM surveys such as DESI \citep{2016arXiv161100036D}, the Vera Rubin Observatory \citep{2019ApJ...873..111I}, the Roman Space Telescope \citep{2015arXiv150303757S}, and CMB-S4 \citep{2016arXiv161002743A}, but also new data collected
with GW standard sirens~\citep{Chen:2017rfc}.
The advantage of standard sirens in this respect is that their measurements present completely different systematics with respect to both CMB and local EM measurements, and will thus deliver completely independent information on the Hubble constant, which will hopefully point towards its correct value.
Similarly, the interesting aspect of the estimates that we obtained on $H_0$ from our analysis is not that we will reach a level of accuracy comparable with current and future EM probes, but that EMRIs detected with LISA will provide yet another complementary measurement of $H_0$, useful to corroborate results from both EM observations and ground-based GW cosmological measurements.
In the likely scenario in which by the time LISA flies we will already have solved the Hubble tension, the measurements of $H_0$ with LISA EMRIs will help consolidate our control over all systematics, including the ones affecting GWs measurements with ground-based interferometers, and our confidence over the results obtained, especially in the case in which physics beyond $\Lambda$CDM is discovered.

We can also compare our results for the dynamical DE cosmological scenario with current measurements and forecasts for the future.
At present, distance measurements taken with SN Ia alone, i.e., not combined with CMB observations, constrain $w_0$ at the $\sim0.2$ ($\sim$20\%) level~\citep{Scolnic:2017caz} (this number is obtained by fitting simultaneously for $\Omega_m$, but not for $w_a$).
Future EM observations collected with the Euclid mission are expected to improve the constraints up to $\Delta w_0 \simeq 0.06$ and $\Delta w_a \simeq 0.26$~\citep{Amendola:2016saw},
while DESI should provide similar results~\citep{2016arXiv161100036D}.
The results we obtained for $w_0$ in Sec.~\ref{sec:results} reach the same level of accuracy of present EM observations in the worst case scenario (M5), while they match the expected results from the Euclid or DESI surveys in the best case scenario (M6).
This implies that cosmological measurements with LISA EMRIs will be of great interest to probe the DE equation of state, not only because they will reach constraints comparable with EM probes, but most importantly because, as stressed above, they will provide this level of accuracy with completely different systematics, increasing our confidence on such measurements (especially if the cosmological constant value $w_0=-1$ appears to be ruled out).

The constraints obtained on the other two parameters considered in our analysis, namely $\Omega_m$ for $\Lambda$CDM and $w_a$ for DE, seem not to be comparable with the level of precision of current and future EM observations~\citep{Scolnic:2017caz,Amendola:2016saw}.
This means that although they will be affected by completely different systematics, most likely they will not be able to provide useful additional information with respect to EM observations.
In other words, the situation will be similar to the $H_0$ constraints that we have today with GWs~\citep{Abbott:2017xzu}: even if they constitute a complementary measurement of $H_0$, affected by completely different systematics, the level of precision of current GW observations does not offer a result comparable to EM measurements, implying that interesting cosmological insight beyond broad consistency between different measurements cannot be obtained.

Better constraints on cosmological parameters beyond $H_0$ and $w_0$ might be obtained by extending our analysis to EMRIs with lower SNR, since ideally we would have additional points to place on our $d_L-z$ plot.
However, as noticed in Sec.~\ref{sec:detecting_emris_with_lisa} and as discussed in App.~\ref{sec:app_low_snr}, the statistical identification method presents some difficulties when applied to high-redshift EMRI events having a large number of potential hosts. 
In our present framework the modelling and inference of galaxy evolution features as well as the inclusion of low-SNR EMRIs, which could allow to better probe further cosmological parameters, are not taken into account and will be investigated in future studies.
A more promising strategy nevertheless may be to combine measurements from different GW sources detected by LISA, as discussed hereafter.

\subsection{Comparison with other GW cosmological measurements}

As we mentioned in Sec.~\ref{sec:introduction}, EMRIs are not the only LISA GW sources that can be used as standard sirens.
Other investigations considering SOBHBs and MBHBs detected by LISA have already produced cosmological estimates.
How do our results compare with those?

LISA SOBHBs will be mainly detected at low redshift ($z\simeq 0.1$) and thus will be only useful to constrain $H_0$.
Recent analyses~\citep{Kyutoku:2016zxn,DelPozzo:2017kme} suggest that SOBHBs could be used to probe $H_0$ at the few \% level, similarly to the results we obtained with EMRIs.
SOBHBs can however reach this level of accuracy only if the detection rate will fall in the optimistic range, which will crucially depend on the yet uncertain sensitivity that LISA will be able to achieve at high frequencies~\citep{Moore:2019pke}.
Note also that an ``archival'' search for SOBHBs will be possible: detecting the black hole merger with the third generation of ground-based GW detectors and then search those sources in the (archival) LISA data with the reduced prior~\citep{Ewing:2020brd}.
On the other hand, EMRIs rely on the LISA sensitivity around the mHz (middle-band) which is guaranteed to be at the level considered in our analysis~\citep{Audley:2017drz}, if not better.
Nevertheless as we have seen, EMRI expected rates are affected by other uncertainties of astrophysical and theoretical nature.
It is important to notice that LISA will fail in deliver compelling low-redshift cosmological results on $H_0$ only if both the LISA high-frequency sensitivity does not perform as expected \textit{and} the true astrophysical population of EMRIs falls on the very pessimistic side, thus not providing enough detections.
In all other scenarios we might expect LISA to deliver useful low-redshift constraints on $H_0$, from either SOBHBs, EMRIs, or both.

LISA MBHB mergers, which are expected to produce observable EM counterparts, will instead probe the expansion of the Universe at much higher redshift, providing in this way complementary constraints to the ones obtained at low redshift with both SOBHBs and EMRIs. LISA MBHBs might yield constraints on $H_0$ at the few \% level~\citep{Tamanini:2016zlh,Tamanini:2016uin,Belgacem:2019pkk}, comparable to SOBHB and EMRI results.
This implies that LISA will deliver accurate measurements of $H_0$ from data sets at different redshift ranges, possibly providing further insights into the current Hubble tension between local measurements (low-$z$) and CMB observations (high-$z$). LISA MBHBs will moreover be useful to probe other cosmological parameters which cannot be constrained at low redshift, for example $\Omega_m$ for $\Lambda$CDM. 
The integration of low- and high-redshift standard sirens will thus allow LISA to test the expansion of the Universe with GWs from early-to-late cosmological times, without the need to combine it with other probes.
This makes LISA a unique cosmological observatory~\citep{Tamanini:2016uin}, whose thorough implications are currently being investigated and will be presented in a future study.

Finally, we can compare our cosmological results with the ones forecast for future Earth-based GW interferometers.
The current LIGO-Virgo network of detectors, with further improvements and the addition of more interferometers, is expected to produce constraints on $H_0$ at the few \% level only if the rate of multi-messenger detections will lie on the optimistic side~\citep{Chen:2017rfc}.
Future third-generation detectors, such as ET~\citep{Punturo:2010zz, Maggiore:2019uih}, are instead expected to precisely probe the expansion of the Universe up to $z\sim2$~\citep{Sathyaprakash:2009xt,Taylor:2012db,DelPozzo:2015bna,Belgacem:2019tbw}.
They will provide complementary results with respect to LISA in approximately the same time period (2030s), implying that cross-checking GW cosmological measurements from space and from the ground will increase our confidence in the obtained results and will help us to get a handle on the expected systematics.

\subsection{More optimistic EMRI scenarios}

Because of the reasons mentioned in Sec.~\ref{sec:detecting_emris_with_lisa}, in our cosmological investigation we selected and analysed only three out of the twelve EMRI models considered in~\citet{Babak:2017tow}.
In first approximation, i.e.,~by considering a similar detected SNR distribution among different EMRI models, we can expect that our results can be extrapolated to the other EMRI models according to the square-root of the number of detected events.
In other words, the constraints we found on $h$ and $w_0$ should improve by a factor inversely proportional to $\sqrt{N}$~\citep{Schutz:1986gp}, where $N$ is the total number of EMRIs useful for cosmology, reported in Sec.~\ref{sec:detecting_emris_with_lisa} for each model.
Note however that different EMRI models will in general provide different SNR distributions, mainly because of their different underlying astrophysical features.
For this reason the following estimates should be considered only as rough predictions, providing only an approximate, order-of-magnitude indication of the results that other EMRI models could possibly offer.

Unfortunately, it is difficult, if not impossible, to estimate the number
of cosmologically useful events
for the other EMRI models of~\citet{Babak:2017tow} without performing the detailed analysis we carried out for the models M1, M5, and M6.
For this reason, in what follows we use the total number of detected EMRI events, as given by Table~I in~\citet{Babak:2017tow} (AKK column), as a rough proxy for this number.
We must keep in mind however that different EMRI scenarios will of course yield a population of EMRIs with different properties, which will certainly translate in a different number of cosmologically useful events after one applies the threshold we defined in Sec.~\ref{sec:detecting_emris_with_lisa}.
We stress however that the
estimates that follow should only be taken as simple indicative numbers useful to quickly explore other possible EMRI scenarios, rather than statistically robust results.

The majority of the EMRI scenarios considered in~\citet{Babak:2017tow} present a number of detections comparable to M1, our fiducial model.
We can thus expect that for all these scenarios we would achieve cosmological constraints comparable to the ones we obtained with M1.
Model M8 yields a number of detections similar to M5, and thus for M8 we expect roughly the same cosmological results as the more pessimistic ones we found in our analysis.
The most pessimistic model of~\citet{Babak:2017tow}, namely model M11, provides only one detected event per year, way below the detections achieved with M5, our pessimistic scenario.
Needless to say, we would not expect any useful cosmological results with M11.

On the other hand, there are three models which predict more detections than those obtained in our optimistic scenario M6.
We can thus extrapolate results starting from the detections of M6 by using our simple $\sqrt{N}$ estimate.
Model M3 presents roughly twice the detections obtained with M6, which translates into improved 4-year (10-year) constraints (90\% C.I.) as 
$\Delta h \simeq 0.0092$ (0.0062) and $\Delta w_0 \simeq 0.052$ (0.045), corresponding to relative errors of 1.3\% (0.84\%) and 5.2\% (4.5\%), respectively.
The two most optimistic scenarios are however models M7 and M12, which deliver an order of magnitude more EMRI detections with respect to M6.
Within M7 we estimate improved 4-year (10-year) constraints as $\Delta h \simeq 0.005$ (0.034) and $\Delta w_0 \simeq 0.029$ (0.025), corresponding to 0.69\% (0.46\%) and 2.9\% (2.5\%) relative uncertainties, respectively.
For M12 instead we find $\Delta h \simeq 0.0040$ (0.0027) and $\Delta w_0 \simeq 0.023$ (0.020), corresponding to 0.55\% (0.37\%) and 2.3\% (2.0\%).

According to these rough estimates, we could reach sub-percent constraints on $H_0$ and two-percent constraints on DE ($w_0$) if the rate of EMRIs detected by LISA is at the most optimistic end of expectations.
From a cosmological perspective these would constitute outstanding results, which even EM probes might not be able to achieve in the near future.
Having said that, we must stress again that these are only simple estimates which rely on extremely optimistic EMRI scenarios and do not take into account all the complexities of the true detected population of EMRIs.

\subsection{Future prospects}

The analysis performed in the present study can be improved in several ways.
First of all, the numbers and properties of the EMRI population detected by LISA could be better characterized by considering enhanced astrophysical modelling and by refining the LISA response. 
In the error estimation we have used the long-wavelength approximation to the response which takes into account the amplitude and the phase (Doppler) modulation of the signal due to LISA's motion around the Sun. This modulation is what gives us the sky position used in this paper. The full response also includes the sky-dependent time delays due to GW propagation across satellites in the constellation. Taking the full response, therefore, could improve the sky localisation for EMRIs with $M< \rm{few}\,\times 10^5\, M_{\odot}$.
In addition,
one could decisively improve the data analysis treatment by starting from better EMRI waveforms, possibly including inputs from self-force calculations, and by performing a full Bayesian parameter estimation with a better model for the instrumental noise.
Although such improvements would require further theoretical developments and costly numerical computations, they would permit to consider full non-Gaussian posteriors in both distance and sky localisation, which will be important in making our analysis more realistic, especially for low-SNR events. 

Another important addition that might be taken into consideration is modelling galaxy time evolution in order to characterize departures from homogeneity and uniformity in the employed galaxy catalog (see Sec.~\ref{sec:gw_cosmology_without_em_counterparts}).
This would be a more realistic representation of the Universe, but the price to pay is the need to perform simultaneous inference of both the cosmological parameters and the properties of the galaxy evolution, which on the one hand is computationally costly and might degrade cosmological results, but on the other hand will provide complementary astrophysical insight.

Another improvement would be to account for incompleteness of the galaxy catalogue. In the current analysis we have ignored this and assumed that out-of-catalogue hosts will be close to hosts that are included in the catalogue. This becomes increasingly problematic for more distant events, where a greater fraction of potential hosts will be missing. Incompleteness can be accounted for in the analysis, for example by weighting galaxies in the catalogue by the number of nearby galaxies that are missing, or by adding an appropriate number of missing galaxies into the assumed redshift distribution~\citep{Abbott:2019yzh}. The impact of such corrections should be explored in the future.

One can also think of using galaxy observational features, such as luminosity, mass, metallicity, and others, to weight galaxies differently within each GW sky localisation region.
Moreover, one could use empirical relations, for example between the mass of the MBH located at the center of the galaxy, which is inferred from EMRI parameter estimation, and the mass or luminosity of galaxies in order to identify more likely host galaxy candidates.
Although these methods would reduce the number of possible host galaxies for each EMRI event, they would also introduce a dependency upon astrophysical modelling into the analysis, possibly introducing new systematics, or the need to marginalise over additional astrophysical parameters.

We conclude this section by noting a few aspects of our inference model that could be improved.
First of all, we decided to neglect the intrinsic EMRI population evolution in the prior assignment over $z_{gw}$. The possibility of some bias in the estimation of the cosmological parameters due to a rapid evolution of the rate of EMRIs with redshift, which means that the weighting within each box should not be uniform in $z$, has been already pointed out in~\citet{MacLeod:2007jd}, although they did not account for this effect in their analysis. Just like for the cosmological evolution of the galaxy population, inclusion of this additional feature might degrade the overall inference over $\Omega$. However, in scenarios  where the number of detected EMRIs is large, this term would impose additional constraints on the whole EMRI population through the time dependence of the merger rate, thus potentially increasing the amount of information contained in each posterior. Moreover, we would be able to relax the arbitrary SNR cutoff of $100$, using more faintest and possibly farther events, thus exploring a larger co-moving volume of the Universe.

Finally, it is worth reporting that preliminary investigations of the full $\Lambda$CDM cosmological model, where the curvature term $\Omega_k$ is not fixed to 0, seem to indicate that, even for moderate-redshift sources such as our loudest EMRIs at $z\lesssim 0.7$, LISA will provide some simultaneous constraints on all cosmological parameters~\citep{laghi2021Moriond}.
In our fiducial scenario, preliminary results indicate that $\Omega_k$ and $\Omega_m$ can be constrained with an accuracy of $\sim30\%$ and $\sim55\%$, respectively, while retaining an accuracy on $H_0$ of $\sim2\%$.
These preliminary results will need to be further investigated in the future; however, they suggest that a possible simultaneous inference of cosmic curvature and other cosmological parameters with LISA standard sirens will indeed be possible.

We stress that the present investigation constitutes a first simple attempt at assessing the cosmological potential of LISA EMRIs.
Further studies, improving the analysis along the lines outlined above, will be needed in order to provide more reliable forecasts and to prepare for the developing of the pipelines needed to analyse the expected data from LISA.


\section{Conclusion} 
\label{sec:conclusion}

In this article we investigated the cosmological potential of EMRIs detected by LISA.
By statistically matching the sky localisation region of the loudest EMRI events (as given by the analysis of \citet{Babak:2017tow}) with the position and redshift of galaxies within a given catalog (in our case based on the Millennium run), we extracted constraints on the parameters characterizing the background evolution of two cosmological models: $\Lambda$CDM and a dynamical DE scenario.
Our results show that interesting cosmological insight can be gained from EMRIs as standard sirens. Over three different EMRI models and two LISA mission durations, constraints on $H_0$ and $w_0$ (the equation of state of DE) can respectively reach the $\sim$1.1\% and $\sim$4.3\% levels in the best case scenario, and be degraded to maximum uncertainties of $\sim$3.6\% and $\sim$12.3\% in the worst case scenario (cf.~Fig.~\ref{fig:summary}). In particular, for our fiducial model M1, EMRI observations are expected to constrain $H_0$ and $w_0$ to $\sim$2.5\% (1.5\%) and $\sim$7.4\% (6.2\%) respectively, assuming a four (ten) year LISA mission.  
As we discussed at length in Sec.~\ref{sec:discussion}, these results will be of great value for cosmology, and will increase our confidence on other EM and GW measurements.

Our results are largely compatible with those reported in \citet{MacLeod:2007jd}, the only EMRI cosmological analysis available in the literature to date.
In that analysis it was claimed that 1\% accuracy on $H_0$ can be reached with as few as 20 EMRIs at redshift $z<0.5$. However, that analysis assumed parameter estimation accuracies appropriate for a more optimistic LISA design. 
It was shown in~\citet{Gair:2017ynp} that 7/1/8 EMRI events satisfying these optimistic parameter constraints were found in two years of observation for models M1/M5/M6, respectively. Therefore, we would expect to achieve comparable precision with ten years of observation of models M1 and M6, as we find here. The fact that we do not do better, even with the slightly larger number of well-localized events and the inclusion of additional less well-localized events in the analysis, is most likely due to the simplifications
employed in the analysis in~\citet{MacLeod:2007jd}, such as the use of a linear cosmic expansion model with $H_0$ as the parameter.

Finally, as in~\citet{MacLeod:2007jd}, we let each EMRI source contribute on an equal footing to the likelihood, while, differently from~\citet{MacLeod:2007jd}, we do not assume equal probability over the whole sky localisation region, but weight galaxies by the marginalised likelihood, see Eqs.~\eqref{eqn:prior_zgw} and~\eqref{eq:margweights}.

As a general final remark, we stress that LISA, a space mission dedicated to GW science, will reveal itself as a unique cosmological probe, through which we will be able to map the expansion of the Universe using different GW sources as standard sirens at different redshifts, including, as thoroughly demonstrated by our study, EMRIs.
Future more-in-depth investigations may deliver cosmological forecast analyses with LISA EMRIs extended at higher redshift, thus allowing us to explore the high-redshift Universe with dark sirens.

\section*{Acknowledgements}

D.L. and W.D.P. thank Stefano Rinaldi for discussions. N.T.~acknowledges partial support from the COST Action CA16104 “Gravitational waves, black holes and fundamental physics” (GWverse), supported by COST (European Cooperation in Science and Technology).
A.S. is supported by the European Research Council (ERC) under the European Union’s Horizon 2020 research and innovation program ERC-2018-COG under grant
agreement No 818691 (B Massive).

\appendix

\section{Discussion of caveats and systematics}
\label{app:appendix}

Here we discuss three aspects of the results presented in the paper: (i) how a lower galaxy mass threshold affects our estimates of the cosmological parameters; (ii) the tendency towards high values of the parameter $w_a$ observed in the 10-year $M6$ scenario; (iii) the consequences of imposing a low cutoff in SNR on the $\Lambda$CDM and DE analyses.

\subsection{Low galaxy mass threshold}
\label{sec:app_gal_mass_thr}

The full-sky galaxy catalog of~\citet{2012MNRAS.421.2904H} is only complete to AB magnitude $i<21.0$. This means that the galaxy mass completeness of the catalog is a function of redshift. The catalog is complete out to $z\approx 0.5$ only for galaxies with $M_*>10^{10}\msun$, which is the main reason we used this threshold for our study.

Most EMRI events detected by LISA, however, involve MBHs with $M\sim10^6\msun$, that might be hosted in galaxies below this mass threshold cut. In fact, $10^6\msun$ MBHs are observed in galaxies spanning a vast stellar mass range, $10^9\msun<M_*<10^{11}\msun$~\citep{2015ApJ...813...82R}. Although the occupation fraction of MBHs in galaxy nuclei starts to decline at $M_*<10^{10}\msun$, it is also true that the galaxy number density increases at low masses ~\citep{2010ApJ...714...25G,2019BAAS...51c..35G}. It is therefore potentially dangerous to set a host mass limit at 
$10^{10}\msun$.
We notice that \citet{Petiteau:2011we} performed a study similar to ours, focusing on MBHBs. In their analysis, they used catalogs with different apparent magnitude cuts, $m_r=24, 25, 26$, finding consistent results. They argued that this is due to the rough self similarity of galaxy clustering. 
Since galaxies tend to accumulate in groups and clusters, small-mass galaxies have clustering properties similar to more massive ones. Therefore, lowering too much the mass (or magnitude) threshold of the hosts included in the analysis might result in a large computational burden without having significant impact on the inference. Clustering self similarity, however, does not hold for all types of galaxies.
In particular, \citet{2021ApJS..252...18T} recently showed that blue low-surface-brightness galaxies are much more evenly distributed than their red counterparts. It is unclear, however, whether an MBH can grow and, most importantly, whether EMRIs can efficiently form in such low density systems. In fact, EMRI rates of $\approx 100$ Gyr$^{-1}$ are expected for galaxies with stellar and compact object mass densities of the order of $10^5-10^6\msun$ pc$^{-1}$ in the central parsec, and those rates likely scale more than linearly with density, due to the inefficiency of relaxation and mass segregation in low density environments \citep[e.g., see][]{2011CQGra..28i4017A}.

Besides all of these considerations, to check that our conclusions do not depend on the specific mass cut used in our work, we performed a tailored simulation employing a more complete galaxy catalog. We used a light cone covering 1/8 of the sky, including all galaxies down to $M_\star = 3 \times 10^{9}\msun$ up to $z=1$. The sky map was constructed using L-Galaxies, a state of the art semi-analytic model implemented on the Millennium run, following the procedure outlined in \citet{2019A&A...631A..82I}. Having the freedom to construct our own sky map, we pushed the mass cut down to the limit allowed by the mass resolution of the Millennium run, which is about $M_\star = 3 \times 10^{9}\msun$. Due to this lower threshold, the number of putative hosts in each error-box is about a factor of five higher. 

\begin{figure}
\includegraphics[width=0.45\textwidth]{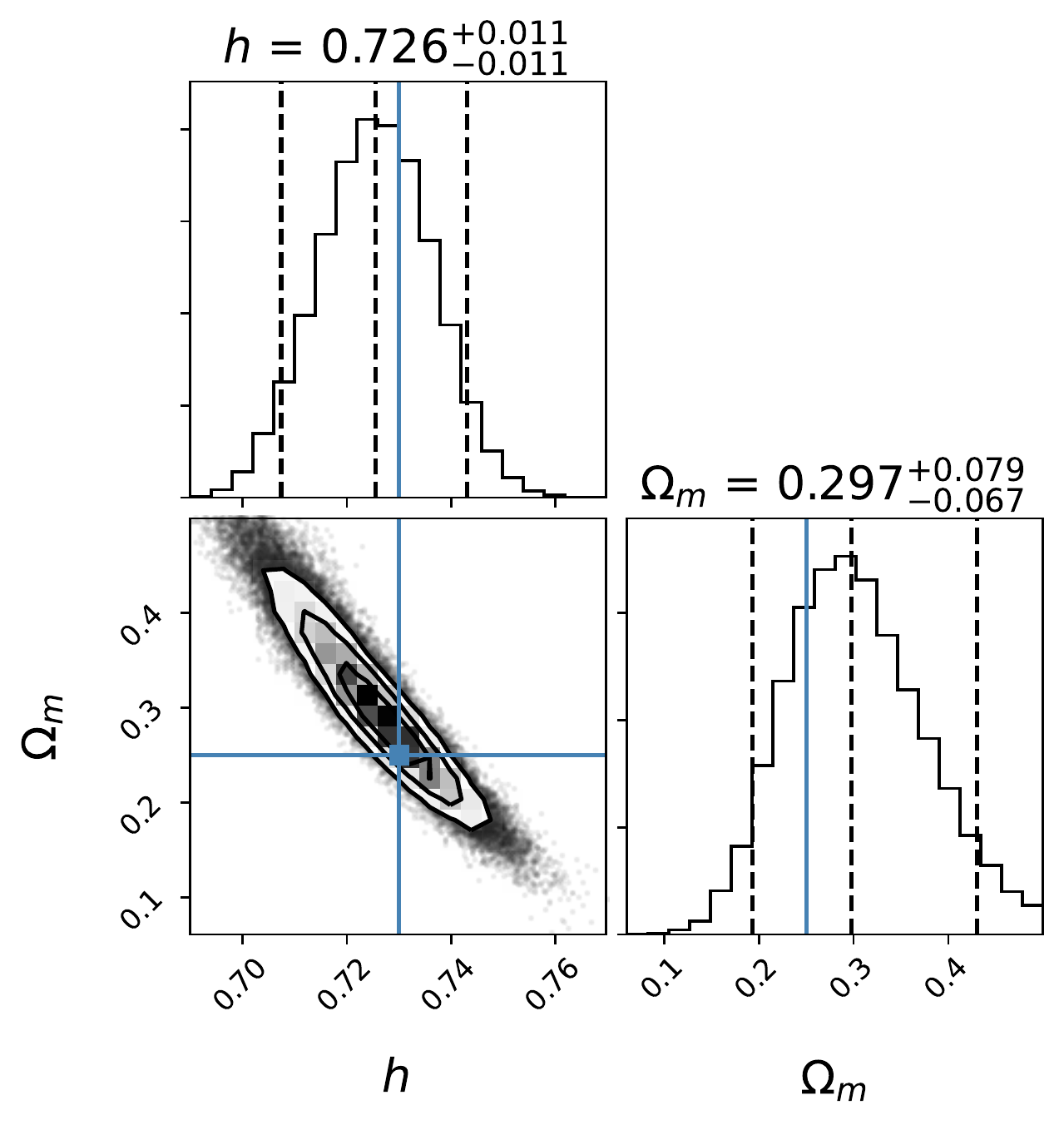}
\caption{Corner plot of the posteriors for the parameters $h$ and $\Omega_m$ in the Lambda CDM scenario for the M1 model (10 years), obtained averaging over six different realisations of the loudest events used in the main paper but with a different light cone having a resolution down to $M_*>3 \times 10^{9}\msun$.}
\label{fig:M1_LambdaCDM_gal_mass_thr}
\end{figure}

Using this new sky map we ran six realizations of model M1 for the $\Lambda$CDM case, assuming 10 years of observations. We employed the same selection criteria, keeping only EMRIs observed at SNR$>100$.
Results are reported in Fig.~\ref{fig:M1_LambdaCDM_gal_mass_thr}, obtained by averaging over the six different realisations. It can be seen that the precision on the measure of $h$ and $\Omega_m$ is comparable to the one reported in Fig.~\ref{fig:corner_lcdm} for the same model, showing that low-mass hosts are not expected to dramatically affect our expectations on the measure of the cosmological parameters. 

\subsection{Increasing $w_a$}
\label{sec:app_railing_wa}

The right panels of Fig.~\ref{fig:corner_de} show the results for $w_0$ and $w_a$ in the 4 and 10-year optimistic scenarios (M6). 
It can be noted that while $w_0$ is correctly measured (being the true value always well within the 90\% credible intervals), $w_a$, although uninformative, shows a railing against the upper prior bound. This is particularly evident in the 10-year analysis (bottom right panel of Fig.~\ref{fig:corner_de}, where the median value is ~40\% off the true value and the correlation between $w_0$ and $w_a$ tends to push for $w_0<-1$. It is worth to focus on this specific scenario and see what is the possible cause of this behaviour. 

\begin{figure}
\includegraphics[width=0.45\textwidth]{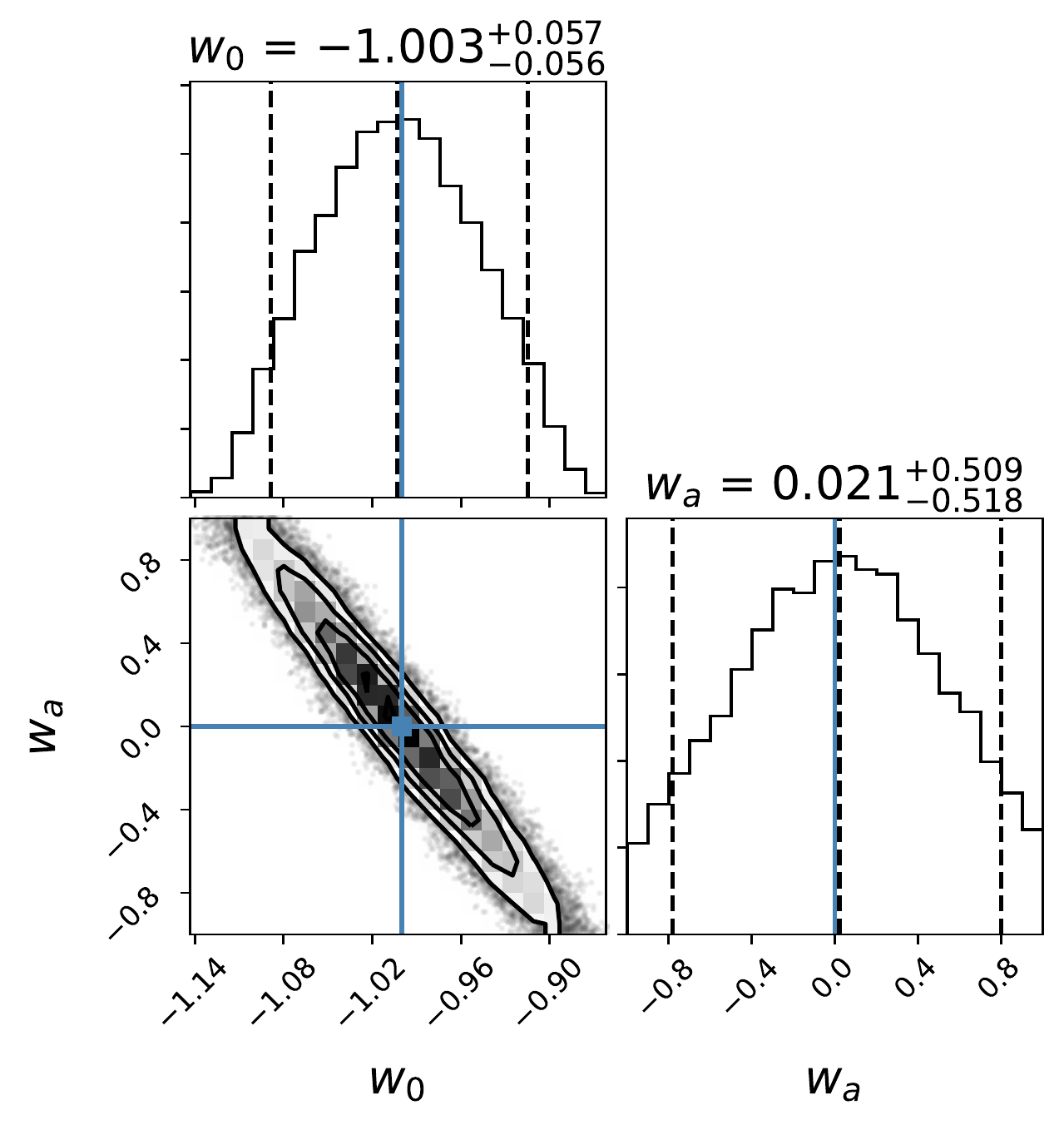}
\caption{Corner plot of the posteriors for the parameters $w_0$ and $w_a$ in the DE scenario for the M6 model (10 years), obtained analysing the same set of events used to produce the analogous corner plot in Fig.~\ref{fig:corner_de}, but now keeping only the nearest-in-$z$ host for each EMRI. As can be seen, the railing behaviour of $w_a$ has now disappeared.}
\label{fig:M6_DE_no_railing}
\end{figure}

We use the same set of 73 loudest EMRI events used in that scenario, this time associating to each event only one galaxy host, chosen to be the nearest in redshift to the EMRI. This analysis may be viewed as representative of the physical scenario in which all our events have an EM counterpart, so it can be used also to investigate what the DE paradigm would predict in that case. In our likelihood, Eq.~\eqref{eqn:single_event_likelihood}, this choice corresponds to assigning $w_j=0$ to all the other hosts.
The scope of this test is to see whether the railing seen for $w_a$ is due to the particular nature of the EMRI catalog or to our formulation of the problem. The results, shown in Fig.~\ref{fig:M6_DE_no_railing}, seem to indicate that indeed this is imputable to the latter.
The railing is now absent, showing posteriors for $w_0$ and $w_a$ that are fully consistent with the fiducial values. Thus, cross-matching the EMRIs with their nearest-in-$z$ hosts gives substantially unbiased posteriors. This seems to point to some limitations on our formulation concerning the way in which the galaxy hosts are treated.

Looking at Fig.~\ref{fig:M6_DE_no_railing}, it is also interesting to note that the CLs on the measure of $w_0$ are comparable to those of the general case  -- where we include all galaxy hosts -- reported in Fig.~\ref{fig:corner_de}: this shows that even if one were to observe EM counterparts to these EMRIs, the inference would not substantially improve.

\subsection{Low-SNR events}
\label{sec:app_low_snr}

Even though we limited our analysis to high-SNR events, it is interesting to estimate the contribution of low-SNR events to the inference of the cosmological parameters under the assumption of likelihood~\eqref{eqn:single_event_likelihood}.
As an illustrative example, let us consider the EMRI catalogs for our fiducial model M1 in the 10-year mission case. We will first adopt our first selection procedure detailed in Sec.~\ref{sec:detecting_emris_with_lisa}, which impose constraints on the measurement precision of the sky position and the luminosity distance (still associating to each EMRI hosts drawn from our galaxy catalog up to $z=1$). Then, to limit ourselves to low-SNR events, we impose an upper cutoff SNR$<$40 and analyse the events assuming our two cosmological models. In case of $\Lambda$CDM, the number of events is 30, while for DE there are 39 events, see Table~\ref{tab:NfaintERMI}.

\begin{figure*}
    \centering
    \begin{tabular}{cc}
        \textbf{$\Lambda$CDM, 10yr (SNR $<$ 40)} & \textbf{DE, 10yr (SNR $<$ 40)}\\
        Model M1  & Model M1 \\
        \includegraphics[width=0.43\textwidth]{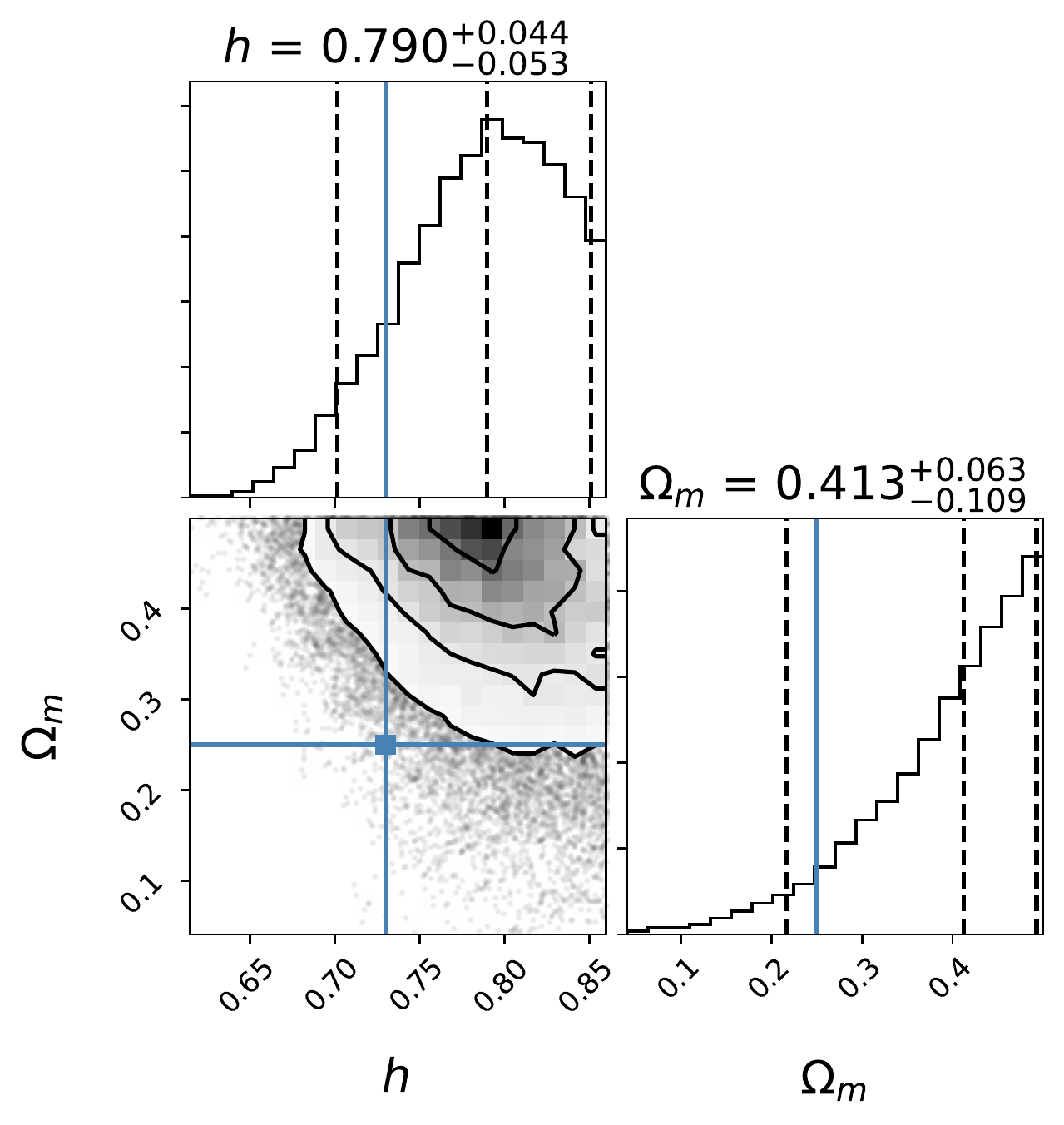}&
        \includegraphics[width=0.43\textwidth]{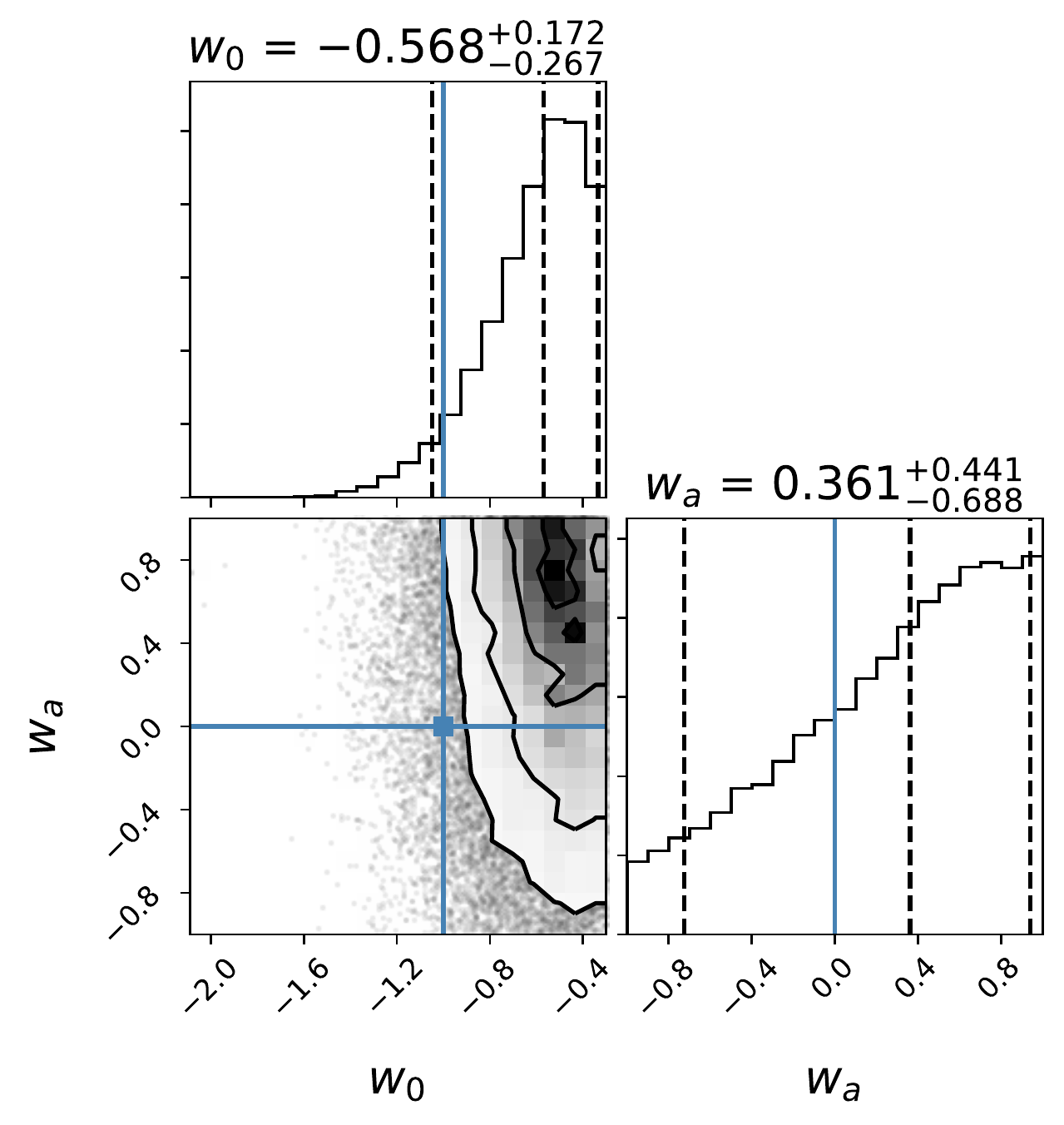}\\
        \textbf{$\Lambda$CDM, 10yr (SNR $<$ 40)}
           &  \textbf{DE, 10yr (SNR $<$ 40)}\\
         Model M1 (nearest host) & Model M1 (nearest host) \\
        \includegraphics[width=0.43\textwidth]{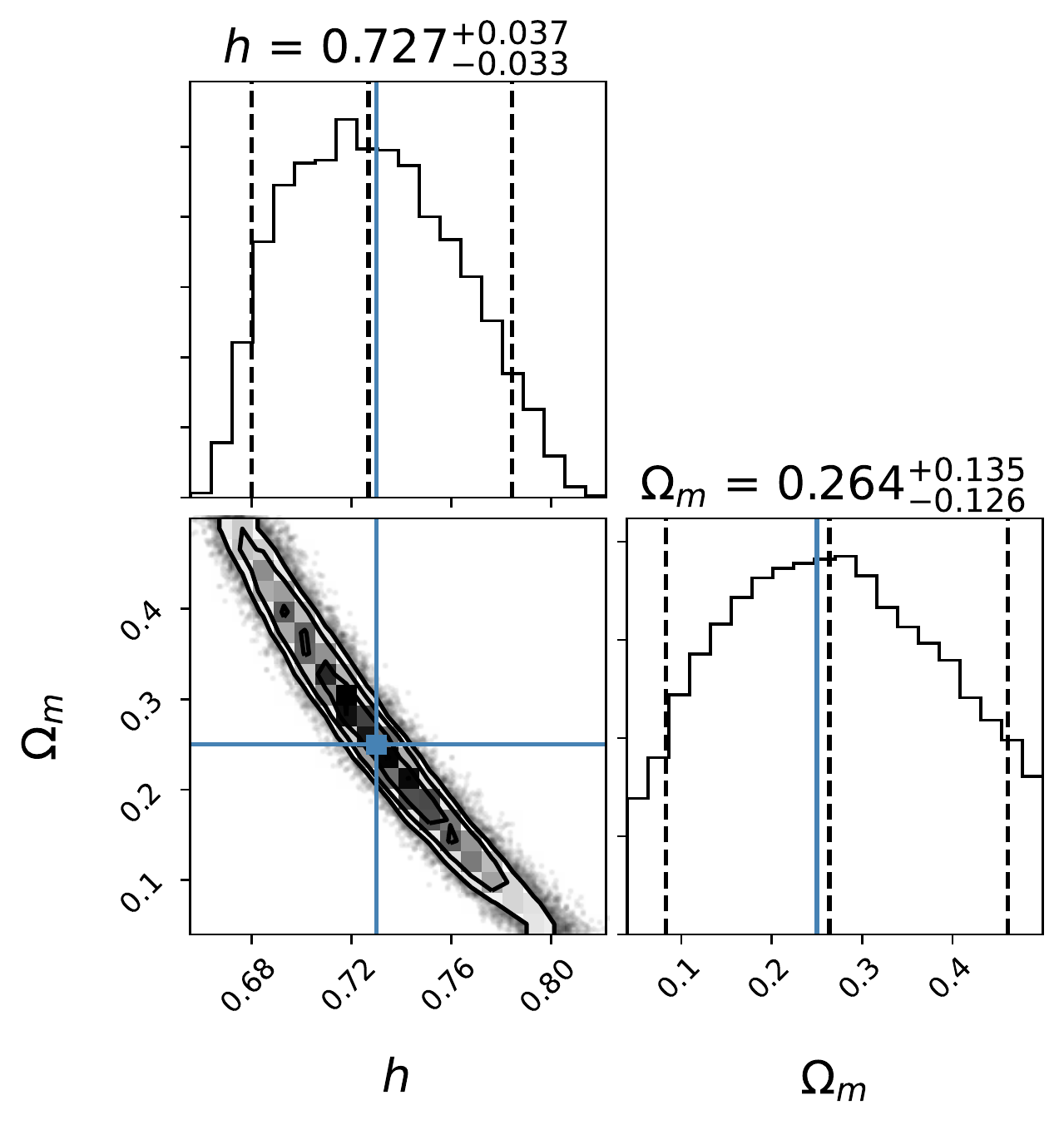}&
        \includegraphics[width=0.43\textwidth]{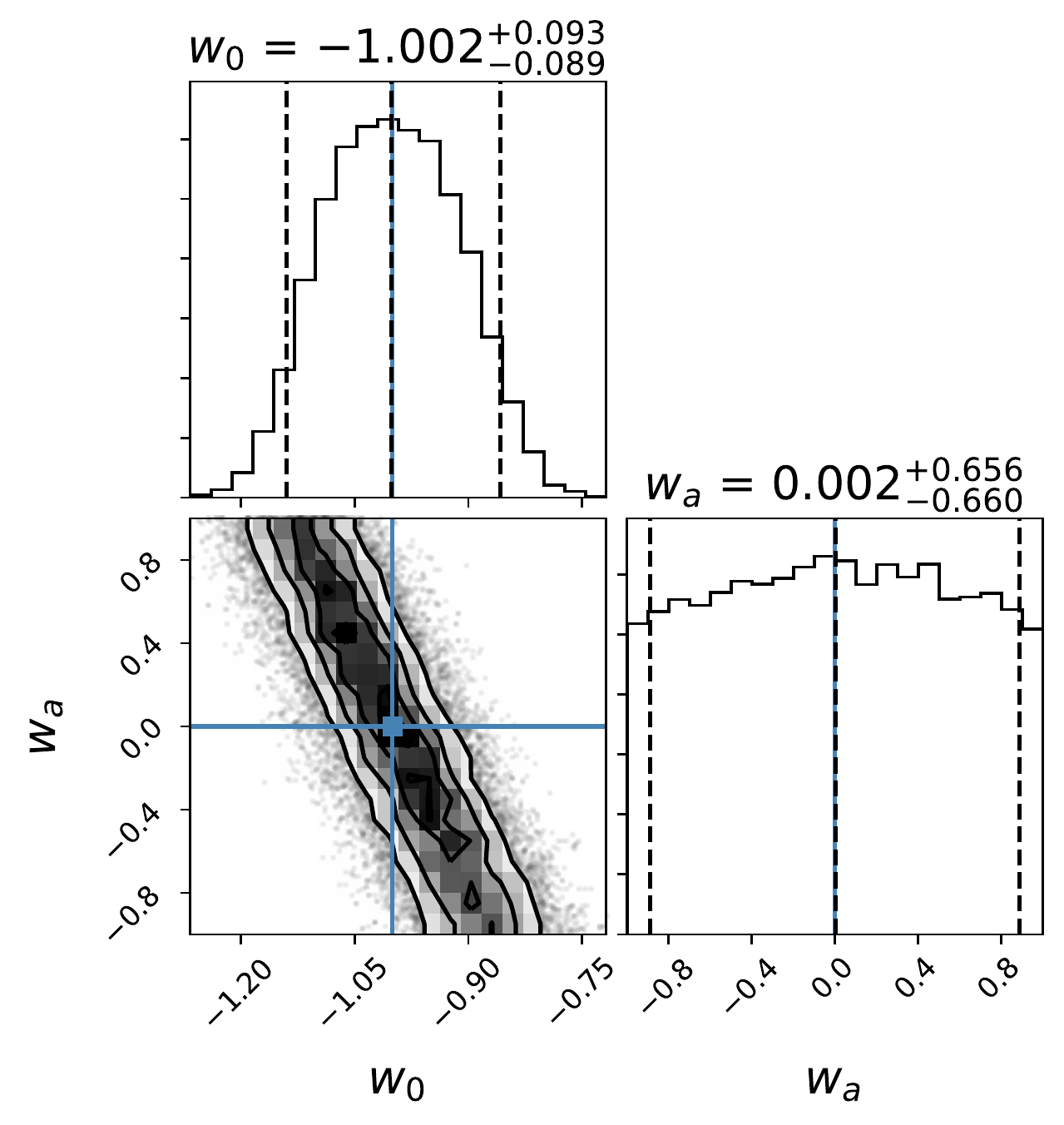}\\
    \end{tabular}
    \caption{Corner plots of the posteriors on the parameters $h$, $\Omega_m$, $w_0$, and $w_a$ in the $\Lambda$CDM and DE scenario, respectively, for our fiducial model M1 (10 years) using the faintest events, selected imposing an upper threshold SNR $< 40$.
    }
        \label{fig:corner_low_snr}
\end{figure*}

The posteriors for the cosmological parameters are shown in the upper-row panels of Fig.~\ref{fig:corner_low_snr}: using only the quietest EMRIs does not yield uninformative posteriors. In particular, the correlation between the two parameters is mainly lost, as expected for low-SNR events, while we observe a preference for high values of the parameters.

To exclude the possibility of sampling issues or that the low-SNR events show some unaccounted-for systematic difference from the rest of the EMRI population, we analysed the same set of low-SNR events, but keeping only the nearest host in redshift to the EMRI, in a manner analogous to what we did in Sec.~\ref{sec:app_railing_wa} for the loudest events of M6 in the DE scenario. 
As already noted in that section, such an analysis may be interpreted as the scenario in which we observe an EM counterpart to all our low-SNR EMRIs.
The resulting posteriors are shown in the bottom-row panels of Fig.~\ref{fig:corner_low_snr}. The correlations between the parameters is restored and the railing has now disappeared, resulting in fairly informative posteriors for $h$ and $w_0$, a mildly informative posterior for $\Omega_m$, and a substantially uniform posterior for $w_a$. 
The posteriors indicate that the low-SNR population of EMRIs is consistent with the general one and that the inclusion of multiple potential hosts in the analysis is likely the culprit. We have thus another indication that our treatment of multiple galaxy hosts is conditioned on our main assumptions, which thus deserve further investigation.
Hence, one should be aware of the caveats to this likelihood in a general analysis where also low-SNR events are included. It is also true that the faintest events are expected to be less relevant to the inference problem, so they should not contribute in a relevant way. This is due to the large number of hosts that are typically associated to each EMRI. The results presented in this work are based on a different selection criterion (see Sec.~\ref{sec:results}) which do not suffer from any relevant bias in the estimate of the main parameters we are interested in, that is, $h$, $\Omega_m$, and $w_0$. 
We plan to further investigate the inclusion of low-SNR events in the cosmological inference elsewhere.

\begin{table}
\centering
\begin{tabular}{c|c|c|c|c|c} 
\hline
\hline
 & \multirowcell{3}{
      detected \\
      (SNR$>$20)
} &  \multicolumn{2}{c|}{$\Delta{d_L}/d_L<0.1$}&  \multicolumn{2}{c}{\multirow{2}{*}{SNR$<$40}}\\
MODEL &  & \multicolumn{2}{c|}{$\&\,\Delta{\Omega}<2\,$deg$^2$}  & \multicolumn{2}{c}{}\\
\cline{3-6}
 &  & $\Lambda CDM$ & $DE$ & $\Lambda CDM$ & $DE$ \\
\hline
\hline
{M1\,({\it fid})} & {2941} & 180 & 202 & 30 & 39\\
\hline
\hline
\end{tabular}
\caption{
Number $N$ of EMRIs observed by LISA in 10 years of operation imposing an upper cut in SNR, as done for the results presented in Fig.~\ref{fig:corner_low_snr}. The upper cutoff SNR$<40$ is imposed in order to study the effect of the faintest events on the inference problem.
\label{tab:faint_events}
}
\label{tab:NfaintERMI}
\end{table}

\section*{Data Availability}

The data underlying this article will be shared on reasonable request to the corresponding author.

\bibliographystyle{mnras}
\bibliography{bibfile}

\bsp	
\label{lastpage}
\end{document}